\documentclass[
onecolumn, notitlepage,
superscriptaddress,
amsmath,amssymb,longbibliography,
aps,floatfix
]{revtex4-1}
\usepackage{graphicx,bm,enumerate}
\usepackage{bm}
\usepackage{color}
\usepackage[breaklinks,colorlinks = true,linkcolor = blue,urlcolor=blue,citecolor=blue]{hyperref}
\usepackage[normalem]{ulem}
\usepackage[dvipsnames]{xcolor}
\graphicspath{{./}{fig/}}

\def \mKt  {\mathcal{K}_{\rm 2}}
\def \mK   {\mathcal{K}}
\def \ww {\bm{w}}
\def \knot {k_{0}}
\def \lamq {\lambda_{\rm q}}
\def \epsl  {\varepsilon_{r}}
\def \hot {\rm{h.o.t}}
\def \deltat {\delta^{\rm 2}}
\def \Sigmap {\Sigma_{\rm p}}
\def \rr   {\bm{r}}
\def \Sdiff {\Sigma_{\rm 2}}

\def \Tnu {\mathcal{D}}
\def \Tp  {\mathcal{T}_{\rm p}}
\def \mm  {{\bm m}}

\def \rr  {\bm{r}}

\def \dab {\delta_{\alpha\beta}}

\def \Cgg {C_{\gamma\gamma}}

\def \EP   {E_{\rm p}}
\def \Bh   {\hat{B}}
\def \mB   {\mathcal{B}}

\def \epT {\varepsilon_{\rm T}}
\def \epsilonf {\varepsilon_{\rm f}}
\def \mulonfp {\mu_{\log\varepsilon_{\rm f/p}}}
\def \siglonfp {\sigma_{\log\varepsilon_{\rm f/p}}}
\def \epsilonp {\varepsilon_{\rm p}}

\def \mD {\mathcal{D}_{\rm f}}

\def \FF     {\bm{F}}

\def \Stwo {S_{\rm 2}}
\def \Sfour {S_{\rm 4}}
\def \Sitwo {\Sigma_{\rm 2}}
\def \Sifour {\Sigma_{\rm 4}}
\def \deltau {\delta u}
\def \Sp    {S_{\rm p}}

\def \tauL {\tau_{\rm L}}

\def \uhat {\hat{\bm{u}}}
\def \mS   {\mathcal{S}}
\def \mup  {\mu_{\rm p}}

\def \ua   {u_{\alpha}}
\def \ub   {u_{\beta}}
\def \ug   {u_{\gamma}}

\def \Sab  {S_{\alpha\beta}}
\def \ab   {\alpha\beta}
\def \muf  {\mu_{\rm f}}
\def \rhof {rho_{\rm f}}
\def \mC  {\mathcal{C}}
\def \Cab {C_{\alpha\beta}}

\def \uhat {\hat{u}}

\def \Stwo {S_{\rm 2}}

\def \rhof {\rho_{\rm f}}

\def \uu  {{\bm u}}
\def \kk  {{\bm k}}

\def  \xx  {{\bm x}}

\def  \ua  {u_{\alpha}}

\def  \ub  {u_{\beta}}

\def \taup {\tau_{\rm p}}
\def \tauf {\tau_{\rm f}}

\def \De  {\mbox{De}}

\def \Rey  {\mbox{Re}}

\def \zetap {\zeta_{\rm p}}

\def \zetat {\zeta_{\rm 2}}

\def \delt {\partial_{t}}
\def \dela {\partial_{\alpha}}
\def \delb {\partial_{\beta}}
\def \delg {\partial_{\gamma}} 
\newcommand{\abs}[1]{\lvert #1 \rvert}
\newcommand{\lrp}[1]{\left( #1 \right)}
\newcommand{\bra}[1]{\left\langle #1\right\rangle}

\newcommand{\eq}[1]{~(\ref{#1})}

\newcommand{\Fig}[1]{Fig.~(\ref{#1})}
\newcommand{\subfig}[2]{Fig.~(\ref{#1}#2)}

\newcommand{\RKS}[1]{\textcolor{blue}{#1}}
\newcommand{\bfig}{\begin{figure}}
\newcommand{\efig}{\end{figure}}
\newcommand{\bc}{\begin{center}}
\newcommand{\ec}{\end{center}}
\newcommand{\bea}{\begin{eqnarray}}
\newcommand{\eea}{\end{eqnarray}}

\newcommand{\oist}{Complex Fluids and Flows Unit, Okinawa Institute of Science and Technology Graduate University, Okinawa 904-0495, Japan}
\newcommand{\tcis}{TIFR Centre for Interdisciplinary Sciences, Tata Institute of Fundamental Research, Gopanpally, Hyderabad 500046, India}
\newcommand{\nordita}{Nordita, KTH Royal Institute of Technology and
Stockholm University, Hannes Alfv\'ens v\"ag 12, 10691 Stockholm, Sweden}

\def \deltap {\delta_{\rm p}}
\def \Xitwo {\xi}

\def \Chit {\chi}
\def \tauK  {\tau_{\rm K}}

\begin{document}
\title{Intermittency in the \textit{not-so-smooth} elastic turbulence}
\author{Rahul K. Singh}
\affiliation{\oist}%
\author{Prasad Perlekar}%
\affiliation{\tcis}%
\author{Dhrubaditya Mitra}%
\affiliation{\nordita}%
\author{Marco E. Rosti}
\email{marco.rosti@oist.jp}
\affiliation{\oist}
\begin{abstract}
Elastic turbulence is the chaotic fluid motion resulting from
elastic instabilities due to the addition of polymers in small concentrations
at very small Reynolds ($\Rey$) numbers.
Our direct numerical simulations show that elastic turbulence, though a
low $\Rey$ phenomenon, has more in common with classical, Newtonian turbulence
than previously thought.
In particular,  we find  power-law spectra for  kinetic energy $E(k) \sim k^{-4}$
and polymeric energy $\EP(k) \sim k^{-3/2}$, 
independent of the Deborah ($\De$) number.
This is further supported by calculation of scale-by-scale energy budget
which shows a balance between the viscous term and the polymeric term
in the momentum equation.
In real space, as expected, the velocity field is smooth, i.e.,
the velocity difference across a length scale $r$, $\delta u \sim r$
but, crucially, with a non-trivial sub-leading contribution $r^{3/2}$
which we extract by using the second difference of velocity.
The structure functions of second difference of velocity up to order
$6$ show clear evidence of intermittency/multifractality. 
We provide additional evidence in support of this intermittent nature
by calculating moments of 
rate of dissipation of kinetic energy averaged over a ball of
  radius $r$, $\epsl$, from which we compute the multifractal spectrum.
\end{abstract}

\maketitle

\section{Introduction}
\begin{figure}
  \centering
  \includegraphics[width=0.45\textwidth]{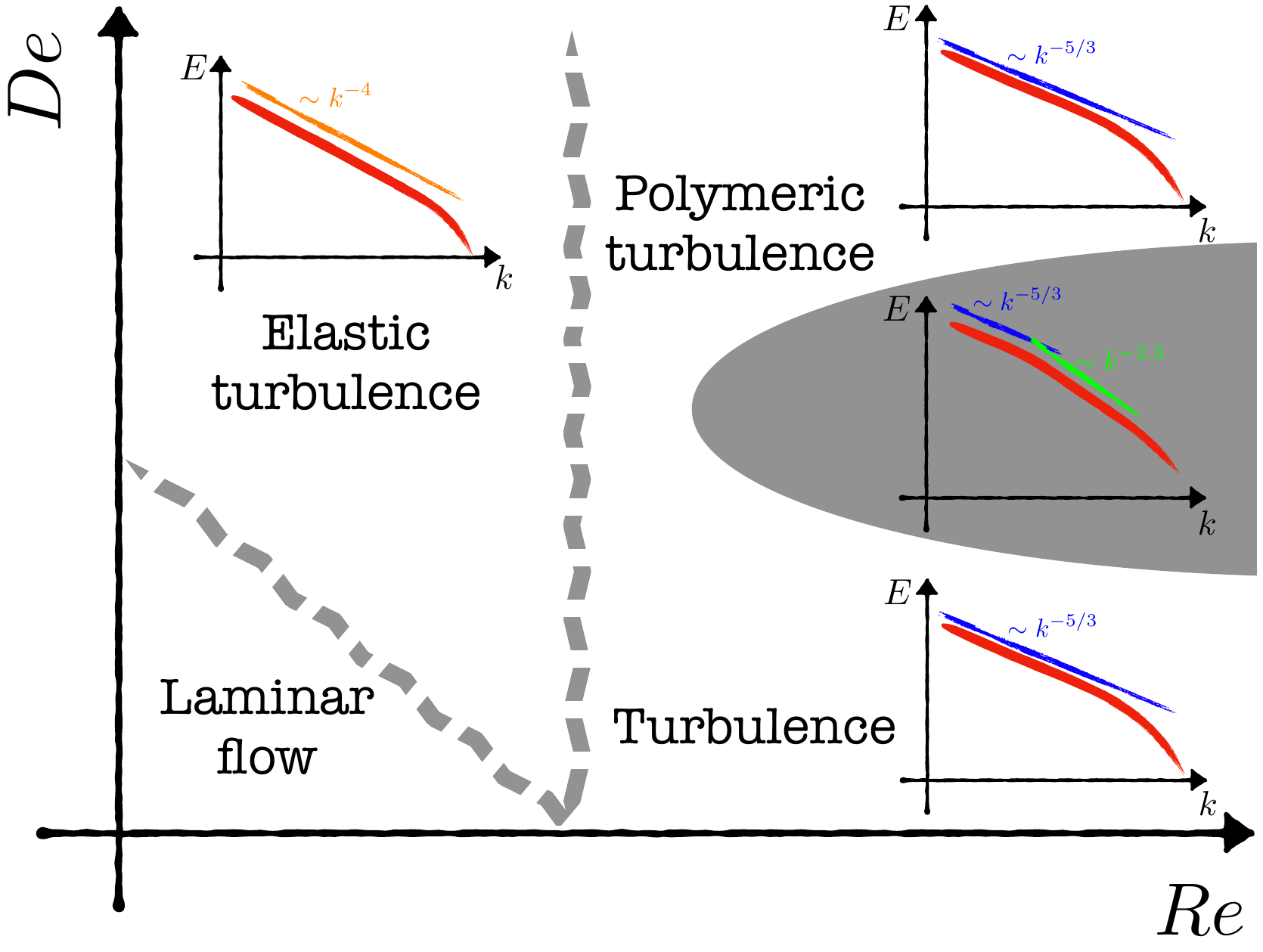}
  \caption{\textbf{Polymeric flows.} An illustrative sketch of the 
    different regimes of polymer-laden flows:
    classical Newtonian \textit{turbulence} (HIT) at large $\Rey$ and zero
    (or small) $\De$, \textit{polymeric turbulence} (PHIT) at large $\Rey$
    and intermediate $\De$, and \textit{elastic turbulence} (ET) at small $\Rey$ and
    large $\De$. The shaded region shows the recently observed elastic scaling regime in addition to the classical Kolmogorov scaling~\citep{Zhang21,Rosti23}.}
  \label{fig:sketch}
\end{figure}
Turbulence is a state of irregular, chaotic and unpredictable fluid motion at very
high Reynolds numbers ($\Rey$), which is the ratio of typical inertial forces over
typical viscous forces in a fluid. 
It remains one of the last unsolved problems in classical physics.
Conceptually, the fundamental problem of turbulence shows up in the
simplest setting of statistically stationary, homogeneous and isotropic
turbulent (HIT) flows: What are the statistical properties of velocity fluctuations?
More precisely, consider the (longitudinal) structure function of
velocity difference across a length-scale $r$ :
\begin{subequations}
  \begin{align}
    \Sp(r) &\equiv \bra{\left[\delta u(\rr)\right]^p} \/,\\
    \text{where}\quad 
    \delta u(r) &\equiv \left[\ua(\xx+\rr) - \ua(\xx)\right]\
    \frac{r_{\alpha}}{\abs{\rr}}\/.
  \end{align}\label{eq:sp}
\end{subequations}
Here, $\uu(\xx)$ is the velocity field as a function of the
coordinates $\xx$ and the symbol $\bra{\cdot}$ denotes averaging over the
statistically stationary state of turbulence.
Here and henceforth we use the Einstein 
summation convention, repeated indices are summed.
The $p$-th order structure function $\Sp$ is the $p$-th moment of
the probability distribution function (PDF) of velocity differences --
if we know $\Sp$ for all $p$  then we know the PDF.
Typically, energy is injected into a turbulent flow at a large length scale $L$,
while viscous effects are important at small length scales $\eta$,
called the Kolmogorov scale, and dissipate away energy from the flow.
In the intermediate range of scales
$\Sp(r) \sim r^{\zetap}$
where scaling exponents $\zetap$ are universal, i.e. they do not depend on how
turbulence is generated.
The dimensional arguments of Kolmogorov give $\zetap = p/3$, which also implies
that the shell--integrated energy spectrum (distribution of kinetic
energy across wavenumbers) $E(k) \sim k^{-5/3} $, where $k$ is the wavenumber. 
Experiments and direct numerical simulations (DNS) over last seventy years
have now firmly established that the $\zetap(p)$
is a nonlinear convex function -- a phenomenon called multiscaling
or \textit{intermittency}.
Even within the Kolmogorov theory, turbulence is non-Gaussian
because the odd order structure functions (odd moments of the PDF
of velocity differences) are not zero.
Intermittency is not merely non-Gaussianity, it implies that
not only a few small order moments but moments of all orders
are important in determining the nature of the PDF. 
We often write $\zetap = p/3 + \deltap$, where $\deltap$ are
corrections due to intermittency. 
A systematic theory that allows us to calculate $\zetap$
starting from the Navier--Stokes equation is the goal of turbulence
research.

Turbulent flows, both in nature and industry, are  often multiphase,
i.e. they are laden with particles, may comprise of fluid mixtures, or
contain additives such as polymers.
Of these, polymeric flows are probably the most curious and intriguing:
the addition of high molecular weight (about $10^{7}$) 
polymers in $10$--$100$ parts per million (ppm) concentration to a turbulent 
pipe flow reduces the friction factor (or the drag) up to 
$5$--$6$ times (depending on concentration)~\citep{Toms77, White08,
  Procaccia08}.
Evidently, this phenomena, called turbulent drag reduction (TDR), 
cannot be studied in  homogeneous and isotropic turbulent flows; nevertheless,
polymer laden homogeneous and isotropic turbulent (PHIT) flows have been 
extensively studied theoretically~\citep{JKB91, JKB96, Fouxon03},
numerically~\citep{Collins2003, Rama03, Rama05, Benzi05, Prasad06, Berti06,
  Schumacher07, Prasad10, Zhang2010, Guido12, watanabe2013hybrid,
  watanabe2014power, Valente14, Bos16,
Valente16, Mani19, Rosti23}, and experimentally~\citep{ Zhang21, 
Schwarz70, McComb77, Tsinober06, Xu09}, 
to understand how the presence of polymers modifies turbulence,
following the pioneering
work by \citet{Lumley73} and \citet{Tabor86}.  
The simplest way to capture the dynamics of polymers in flows
is to model the polymers as two beads connected by an overdamped
spring with a characteristic time scale $\taup$.
A straightforward parameterization of the importance of elastic
effects is the Deborah number $\De \equiv \taup/\tauf$, where
$\tauf$ is some typical time scale of the flow.
In turbulent flows, such a definition becomes ambiguous because
turbulent flows do not have a unique time scale, rather
we can associate an infinite number of time scales
even with a single length
scale~\citep{lvo+pod+pro97,mit+pan04, ray+mit+per+pan11}. 
In such cases, a typical timescale used to define $\De$ is the
large eddy turnover time of the flow, $\tauL$ \citep{Rosti23}. 
The phenomena of PHIT appear at high Reynolds and high Deborah numbers. 

Research in polymeric flows turned into a novel direction when
it was realized that even otherwise laminar flows may become unstable
due to the instabilities driven by the elasticity of
polymers~\citep{Larson92,Shafqeh96}.
Even more dramatic is the phenomena of
\textit{elastic turbulence} (ET)~\citep{steinberg2021elastic},
where polymeric flows at low Reynolds but
high Deborah numbers are chaotic
and mixing, with a shell--integrated kinetic
energy spectrum $E(k) \sim k^{-\Xitwo}$.
It is still unclear whether this exponent is universal or not
-- experiments and DNS in two dimensions have
obtained $ 3 \leq \Xitwo \leq 4 $, and
theory~\citep{fux03} sets a lower bound with $\Xitwo > 3$. 
A three-dimensional DNS of decaying homogeneous, 
  isotropic turbulence with polymers additives (modelled as discrete dumbbells)
  also revealed an exponent $\Xitwo \approx 4$  at late times 
  (with a mild $\De$ dependence), when turbulence had sufficiently decayed and
  elastic stresses were dominant, likely marking the onset of
  ET~\citep{watanabe2014power}.
In summary, as shown in \Fig{fig:sketch}, HIT (in Newtonian turbulence) appears at large $\Rey$ and zero (and small) $\De$;
PHIT appears at large $\Rey$ and intermediate $\De$ number, while ET appears at small Reynolds and large Deborah numbers.

Recently, experiments~\citep{Zhang21} and DNS~\citep{Rosti23}
revealed an intriguing aspect of PHIT: The energy spectrum showed not one
but two scaling ranges, a Kolmogorov-like inertial range at
moderate wave numbers and a second scaling range with $E(k) \sim k^{-2.3}$
resulting purely due to the elasticity of polymers~\citep{Rosti23}.
This is illustrated in the gray shaded region in \Fig{fig:sketch}.
Even more surprising is the observation that both of these
ranges have intermittency correction $\deltap$ which are
the same.
This hints that even at low $\Rey$, where elastic turbulence (ET) appears,
intermittent behavior may exist.
In this paper, based on large resolution DNS of polymeric flows at low Reynolds
number, we show that this is indeed the case.

\section{Model}
\begin{figure}
  \centering
  \includegraphics[width=0.75\textwidth]{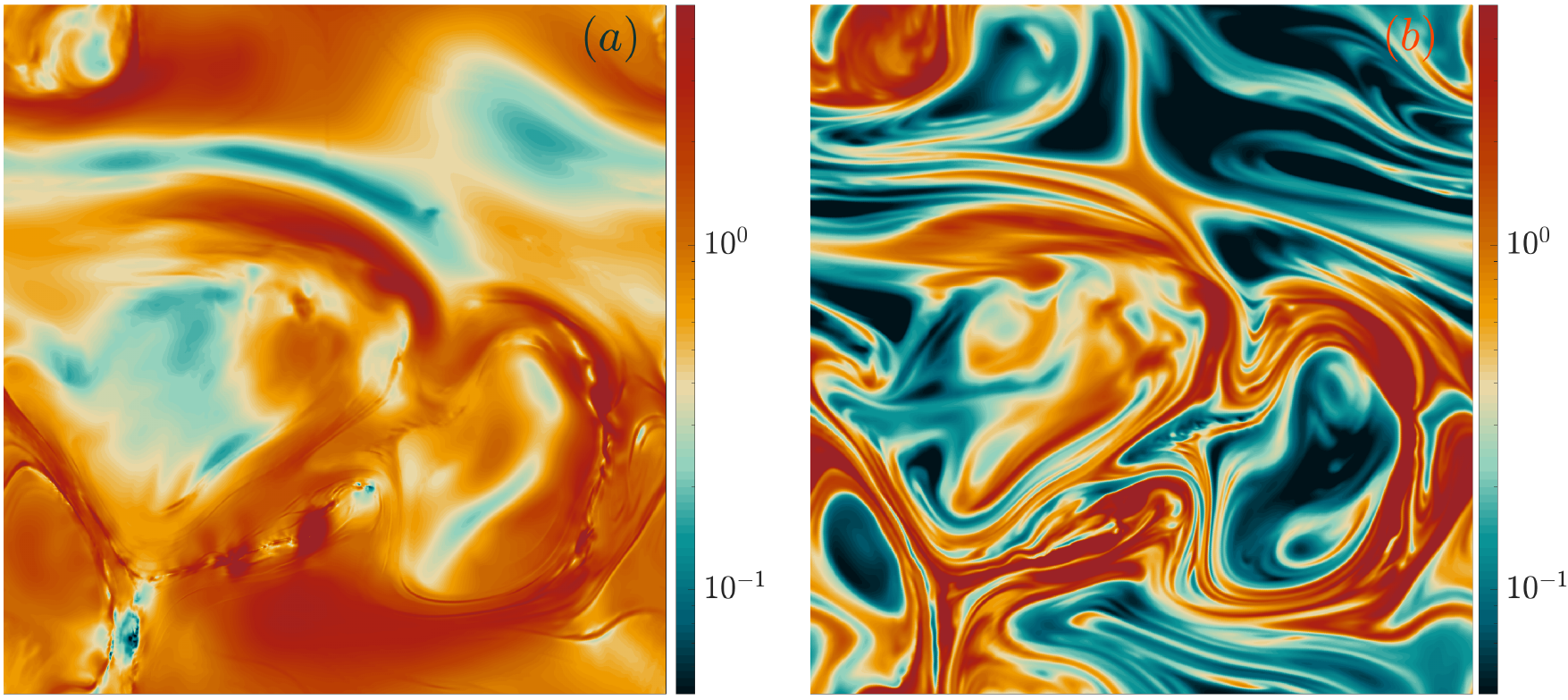}
  \caption{\textbf{Flow visualizations.} 
    Two-dimensional slices of the three-dimensional domain showing snapshots of
    the normalized
    (a) fluid dissipation field $\epsilonf/\bra{\epsilonf}$ and of
    (b) the polymer dissipation field
      $\epsilonp/\bra{\epsilonp}$ in ET for $\De = 9$.}
  \label{fig:snaps}
\end{figure}

We generate a statistically stationary, homogeneous, isotropic flow of a dilute
polymer solution by the DNS of the Navier-Stokes equations 
coupled to the evolution of polymers described by the Oldroyd-B model:
\begin{subequations}
  \begin{align}
    \rhof\left( \delt \ua + \ub \delb \ua \right) &=
    -\dela p +
    \delb \left( 2\muf\Sab  + \frac{\mup}{\taup} \Cab\right) +
    \rhof F_{\alpha}, \label{NN}  \\ 
    \delt\Cab + \ug\delg\Cab &= C_{\alpha\gamma}\delg \ub + C_{\gamma\beta}\dela\ug
    - \frac{1}{\taup}\left(\Cab - \dab\right)  \/.\label{Conf} 
  \end{align}
\end{subequations}
 Here, $\uu$ is the incompressible solvent velocity field, i.e.
$\delb \ub = 0$,
$p$ is the pressure,
$\mS$ is the rate-of-strain tensor with components
$\Sab \equiv (\dela\ub + \delb\ua)/2$,
$\muf$ and $\mup$ are the fluid and polymer viscosities,
$\rhof$ is the density of the solvent fluid,
$\taup$ is the polymer relaxation time, and
$\mC$ is the polymer conformation tensor whose trace $\Cgg$
is the total end-to-end squared length of the polymer.
To maintain a stationary state, we inject energy into the flow using an
Arnold-Beltrami-Childress (ABC) forcing, i.e., 
$\FF = (\muf/\rhof) [ (A \sin z + C \cos y) \, {\bf \hat{x}}
  +  (B \sin x + A \cos z) \,{\bf \hat{y}}
  +( C \sin y + B \cos x) \, {\bf \hat{z}}]$.
The injected energy is ultimately dissipated away by both
the Newtonian solvent ($\epsilonf$) and polymers ($\epsilonp$).
The total energy dissipation rate, $\bra{\epT}$, is given by:
\begin{subequations}
  \begin{align}
  \bra{\epT} \equiv &\bra{\epsilonf} + \bra{\epsilonp} \\
  \text{where}\quad  
  \epsilonf \equiv \frac{2\muf}{\rhof}\left( \Sab\Sab \right) \,; & \;
  \epsilonp \equiv \frac{\mup}{2\rhof\taup^2} \left(\Cgg -3 \right).
  \end{align}
  \label{Diss} 
\end{subequations}
We show typical snapshots of the two energy dissipation rates on
two dimensional slices of our three dimensional DNS in \Fig{fig:snaps}.
Details on numerical schemes and simulations are discussed in the Methods
section.

The Newtonian ABC flow shows Lagrangian chaos in the sense
that the trajectories of tracer particles advected by such a flow
have sensitive dependence on initial condition~\citep{dombre1986chaotic}.
Hence we expect that a polymer advected by the flow will go through a
coil-stretch transition for large enough $\taup$.
The back reaction from such polymers may give rise to elastic turbulence.
The energy spectra of the Newtonian
flows (and those for our non-Newtonian flows with small $\taup$) do not show any power-law
range, and drops-off rapidly in wavenumber $k$, 
see Fig.~(S1a) in the Supplementary Material.
Beyond a certain value of $\taup$, the flow becomes chaotic,
and the resulting flows with $\De \gtrsim 1$ are able to sustain elastic
turbulence.
Henceforth we focus only on the flows that show ET.

Note an important difference between ET and usual Newtonian HIT.
In the latter, the Kolmogorov length (or time) scale is defined
as the scale where the inertial and viscous effects
balance each other.
Although we continue to use the same definition
-- $\Rey_{\lambda}$ and $\tauK$ are calculated from the Newtonian
DNS -- these scales lose their usual meaning because ET appears at
small $\Rey$ at scales where the inertial term is negligible.
The $\eta$ we obtain is, as expected, quite close to the
scale of energy injection.
Therefore, we use the box-size $L$
which is also the scale of energy injection,
as our characteristic length scale.

\section{Results}
\begin{figure*}
 \centering
 \includegraphics[width=.95\textwidth]{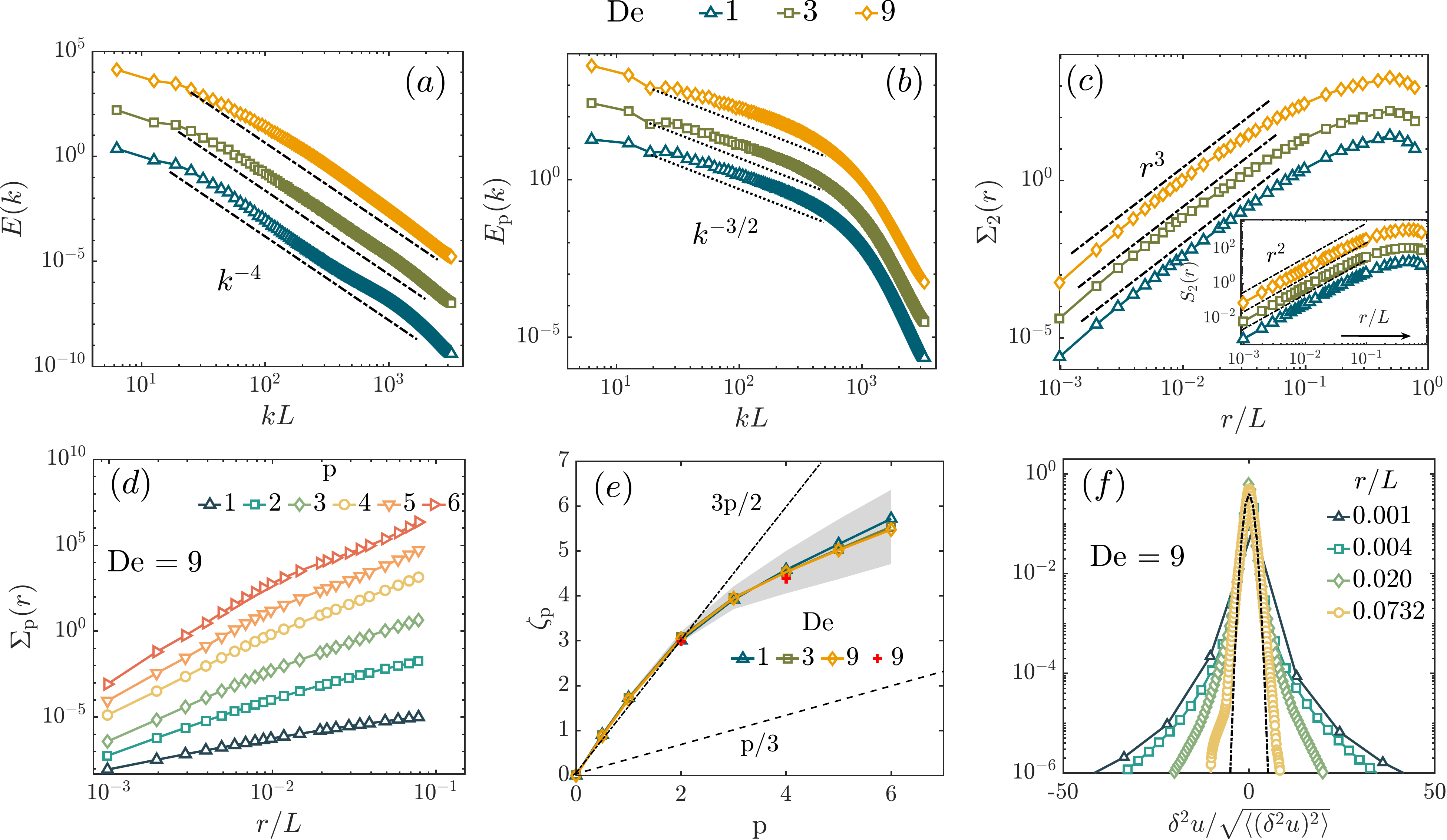}
 \caption{\textbf{Spectra and structure functions.}
   (a) The fluid energy spectra show a universal scaling $E(k) \sim k^{-4}$ independent of $\De$.
    A steeper than $k^{-3}$ fall-off of spectrum means the velocity fields are smooth;
    $\Stwo(r) \sim r^2$ for small $r$, shown in the inset of panel (c).
    (b) The polymer spectra $\EP(k) \sim k^{-3/2}$ follows from the scaling
    of  $E(k)$.
   (c) Plot of the second order structure function of  second differences which scale as $\Sitwo(r) \sim r^3$.
    The exponent is same for the different $\De$, although the range of
    scaling depends weakly on $\De$.
    The inset shows the analytic scaling of $\Stwo(r)$.
    (d) Structure function of second differences, $\Sigmap$, for various orders
    $p$ for $\De = 9$.
    (e) The exponents  $\zetap$ versus $p$, calculated from the scaling
    behaviour of $\Sigmap$.
    Departure from the straight line $\zetap = 3p/2$ shows intermittency.
    Shaded region shows the standard deviation on the exponents
    computed from $18$ snapshots.
    The two red $+$ symbols mark the exponents $\zetat$ and $\zeta_{\rm 4}$
    obtained with an alternate forcing scheme~\citep{esw88}.
    That they lie well within error bars goes on to show that the results are
    independent of the large scale forcing.
 (f) Probability distribution function  of $\deltat u(r)$ for four different
    values of $r$ at $\De = 9$. 
    The distributions are non-Gaussian at small separations, while they become
    closer to a Gaussian (shown as a black dash-dotted curve) for large $r$.
    The corresponding cumulative distribution functions,
    computed using the rank-order method,  are shown in the Supplementary
    Material, section C.}
    \label{fig:Scalings}
\end{figure*}

We present our results for three different Deborah number flows with
$\De = 1, 3$, and  $9$ and  Taylor scale Reynolds number
$\Rey_\lambda \approx 40$.
Let us begin by looking at the (shell--integrated) fluid energy spectrum
\begin{equation}
  E(k) \equiv \int d^3\mm \bra{\uhat(\mm)\uhat(-\mm)}
  \delta (\lvert \mm \rvert - k)\/, \label{eq:Ek}
\end{equation}
where $\uhat(\mm)$ is the Fourier transform of the velocity field $\uu(\xx)$.
We show the spectra for the  three $\De$ numbers in~\subfig{fig:Scalings}{a}.
The spectra $E(k)$ show  power-law scaling  over almost two decades
when plotted on a log-log scale.
Clearly, $E(k) \sim k^{-\Xitwo}$ with $\Xitwo = 4$ independent of the
Deborah number.
Note that in DNS of decaying PHIT, $\Xitwo$ goes from 
$2.3$ to $4$ (and beyond as turbulence decayed) as time
progressed~\citep{watanabe2014power}.
While $\Xitwo = 2.3$ was recently confirmed for PHIT via both
DNS~\citep{Rosti23} and experiments~\citep{Zhang21}, we now show via DNS that
ET is, in fact, a stationary state marked by $\Xitwo = 4$ which is sustained
by purely elastic effects, for a large enough polymer elasticity.

We have verified, using representative DNSs,  that
the scaling exponents of ET remains the
same if we use the FENE-P model for polymers
or a different forcing scheme~\citep{esw88},
that is not white--in--time. 
We have also checked that reducing the resolution to $N^3 = 512^3$
reproduces the same spectra.
Finally, we have turned off the advective nonlinearity in~\eq{NN}
and also obtained the same spectra, thereby confirming
that the turbulence we obtain is purely due to elastic effects. 
We plot all these spectra in
Fig.~(S1) of the Supplementary Material.

We also define the energy spectrum associated with polymer degrees of freedom
as 
\begin{equation}
  \EP(k) \equiv \left(\frac{\mup}{\rhof \taup}\right)\int  d^3\mm\bra{
    \Bh_{\gamma\beta}({\mm})\Bh_{\beta\gamma}(-\mm)}\delta (\lvert\mm\rvert - k)\/,
\label{eq:Ep}
\end{equation}
where  the matrix $\mB$ with components $B_{\alpha\gamma}$ is the (unique)
positive symmetric square root of the matrix $\mC$, defined by
$\Cab = B_{\alpha\gamma}B_{\gamma\beta}$ \cite{balci2011symmetric,nguyen2016small}.
We obtain $\EP(k) \sim k^{-\Chit}$ with $\Chit = 3/2$
as shown in  log-log plot of $\EP(k)$ in~\subfig{fig:Scalings}{b}.
Note that the scaling range of $\EP(k)$ is somewhat smaller than that of
$E(k)$. 
In the statistically stationary state of ET the effect of the
advective nonlinearity must be subdominant.
Hence, at scales smaller than the scale of the external force,
the viscous term in the momentum equation must
balance the elastic contribution~\cite{fux03}.
Using a straightforward scaling argument,
described in detail in the Supplementary Material, section A2 we obtain :
\begin{equation}
  \Xitwo = 2\Chit + 1\/,
\label{eq:XiChi}
\end{equation}
which is satisfied by the values of $\Xitwo$ and
$\Chit$ we obtain.
For further confirmation we calculate all the contributions to the
scale-by-scale kinetic energy budget in Fourier space
(see the Supplementary Material, section A1).
As expected, the contribution from the advective term in~\eq{NN} is
negligible.
Earlier theoretical arguments~\citep{fux03} have suggested
$\Xitwo > 3$ which has also been observed in
experiments~\citep{gro+ste00, gro+ste04, varshney2019elastic}
-- $\Xitwo \approx 3.5$ over less than a decade of scaling range.
We obtain $\Xitwo \approx 4$ which satisfies the
inequality and agrees with shell-model simulations~\citep{SSR16}.
Earlier theoretical arguments~\citep{fux03, steinberg2019scaling} had
also assumed the same balance in the momentum equation that we have,
but in addition had assumed scale separation and a large scale alignment
of polymers in analogy with magnetohydrodynamics, obtaining 
$\Xitwo = \Chit + 2 $, which is not satisfied by our DNS.
  
Note further that experiments often obtain power-spectrum as
a function of frequency and they can be compared with power-spectrum
as a function of wavenumber (typically obtained by DNS) by
using the Taylor ``frozen-flow'' hypothesis~\citep{Fri96}.
In the absence of a mean flow and negligible contribution
from the advective term it is not \textit{a priori} obvious that
the Taylor hypothesis should apply to ET.
We have confirmed from our DNS that a frequency dependent
power-spectrum obtained from time-series of velocity
at a single Eulerian point also gives $\Xitwo = 4$
(see the Supplementary Material, Fig.~(S1b)). 
  
\subsection{Second order structure function}
Next, we consider the second order structure function, $\Stwo(r)$, which is the
inverse Fourier transform of $E(k)$.
This requires some care.
As a background, let us first consider the case of HIT (Newtonian
homogeneous and isotropic turbulence).
Let us ignore intermittency and concentrate on scaling a-la Kolmogorov. 
The second order structure function
is expected to have the following form
\begin{equation}
  \Stwo(r) \sim
  \begin{cases}
     &r^{\zetat} \quad\text{for}\quad  L > r > \eta, \\
     &r^2   \quad\text{for}\quad  \eta > r >0.
  \end{cases}
  \label{eq:StwoK41}
\end{equation}
The range of scales $L > r > \eta$ is the inertial range.
Let us remind the reader that the behaviour $\Stwo \sim r^2$ for
small enough $r$ follows from the assumption that the velocities
are analytic functions of coordinates, which must always hold for any
finite viscosity, however small.
We  call $\Stwo \sim r^2$ the \textit{trivial} scaling.
The strategy to extract the exponent $\zetat$ from DNS is to
run simulations at higher and higher Reynolds number,
which means smaller and smaller $\eta$ to obtain
a significant inertial range from which 
$\zetat$ can be extracted.
In Kolmogorov theory $\zetat = 2/3$, we call this the
\textit{non--trivial} scaling. 
The theory of ET is much less developed than that of HIT.
Nevertheless, we may assume that the velocities must still be
analytic functions. Hence for ET the following must hold
\begin{equation}
  \Stwo(r) \sim r^2 \quad{\rm for}\quad r\to 0\/.
\end{equation}
As this scaling follows directly as a consequence of analyticity
of velocities, we again call this the trivial scaling.
If there is a non--trivial scaling of second order structure
function in ET -- here we are not talking about intermittency
corrections to a non-trivial scaling
but just the existence
of the non-trivial scaling exponent -- it may show up
in the following manner
\begin{equation}
\Stwo(r) \sim
\begin{cases}
   r^{\zetat} \quad\text{for}\quad  L > r > \ell, \\
   r^2   \quad\text{for}\quad   \ell > r >0.
\end{cases}
\label{eq:StwoET}
\end{equation}
This requires introduction of a new length scale $\ell$
which cannot depend on Reynolds number because
in ET we are already in the range of small and fixed Reynolds
number.
The scale $\ell$ may depend on the Deborah number.
To check if it does, we plot $\Stwo(r)/r^2$ for three different
$\De$ ranging from $1$ to $9$ in the Supplementary Material Fig.~(S3b).
At small enough $r$ they all show $\Stwo \sim r^2$.
As $r$ increases they all depart from this trivial scaling
at a length scale $\ell$ which depends very weakly on $\De$,
if at all. 
This implies that even if a non--trivial scaling for $\Stwo$ exists in ET,
it may require DNS at impossibly high $\De$ to
be able to extract $\zetat$. 
Nevertheless, we have now demonstrated that in ET:
\begin{enumerate}[(a)]
\item $\Stwo$ shows trivial scaling at small $r$, i.e.,
  the velocity field is analytic, and 
\item there is departure from the trivial scaling. 
\end{enumerate}
Does the departure from the trivial scaling show a new scaling
range?
To explore this possibility we plot on a log-log scale $\Stwo(r)$
as a function of $r$ in the Fig.~(S3a) of the Supplementary Material, but
it is unclear if there is a clear scaling range at intermediate $r$.
Even if there is a non--trivial scaling exponent, it cannot be
detected from the data, which is the highest resolution DNS
of ET done so far.

It often helps to detect a scaling range if we know beforehand
what the scaling exponent is.
In ET, unlike HIT,  there is no theory that tells us what
$\zetat$ should be, but we do know that the Fourier spectrum
of energy behaves like
$E(k) \sim k^{-\Xitwo}$,
with $\Xitwo \approx 4$ (see~\subfig{fig:Scalings}{a}).
Usual straightforward power
counting implies that 
$\Stwo(r) \sim r^{\Xitwo-1}\sim r^3$ (see section A2 of Supplementary Material for details),
while we obtain $\Stwo \sim r^2$.
This paradox is resolved by noting that,
in the limit $r\to 0$, $r^3$ is subdominant to $r^2$,
hence $\Stwo(r) \sim r^2$ as $r\to 0$ for any velocity field whose spectra
  $E(k) \sim k^{-\Xitwo}$ with $\Xitwo > 3$,
see e.g., Ref.~\citep[][Appendix G]{Pop00}.
This is also known from the direct cascade regime of two-dimensional turbulence
(with Ekman friction) where $E(k) \sim k^{-\gamma}$ with
$\gamma > 3 $ and $\Stwo(r) \sim r^2$,
see e.g., Refs.~\citep[][page 432]{perlekar2009statistically,bof+eck12}.
This suggests that $\Stwo(r)$ satisfies \eq{eq:StwoET} with
$\zetat = \Xitwo-1 \approx 3$.
To test this we plot the compensated second order structure function
$\Stwo(r)/r^3$ as a function of $r$ in the Supplementary Material Fig.~(S3c).
We detect no range at small or intermediate $r$ where this
non-trivial scaling holds.

Now we consider the possibility that 
$\Stwo(r) \sim Ar^2 + B r^3 + \hot$. 
Here, the symbol $\hot$ denotes higher order terms in $r$.
We use a trick~\citep{Biferale01,Biferale03} to extract the
subleading term which scales with the non-trivial scaling exponent:
the idea is to remove the analytic
contribution by considering the \textit{second difference} of velocities:
\begin{subequations}
   \begin{align}
     \deltat u(r) &\equiv \left[\ua(\xx+\rr) -2\ua(\xx) + 
       \ua(\xx-\rr)\right] \left(\frac{r_\alpha}{r}\right)\/,
      \label{eq:d2u}\\
  \text{and define}\quad \Sdiff(r) &\equiv \bra{(\deltat u)^2}.
  \label{eq:Sdiff}
\end{align}
\end{subequations}
 We plot $\Sdiff(r)$ in \subfig{fig:Scalings}{c}. 
We find that $\Sdiff(r)$ shows a significant scaling range
as $r\to 0$ with
the non-trivial scaling exponent $\zetat \approx 3$.
This implies that, 
in ET the velocity fluctuations
across a length scale $r$ can be 
expanded in an asymptotic series in $r$ as
\begin{equation}
    \bra{\delta \ua(\rr)} \equiv
    \bra{\ua(\xx+\rr) -\ua(\xx)}
    \sim G_{\ab}r_{\beta} + H_{\ab}r_{\beta}^h + \hot.,
\label{eq:uexpand}
\end{equation}
where $G_{\ab}$ and $H_{\ab}$ are (undetermined) expansion coefficients, and $h \approx \zetat/2 = 3/2$.
The use of $\Sdiff$ is necessary to extract the subleading contribution.

To appreciate the importance of this result, let us revisit the Kolmogorov
theory of turbulence:
in  the limit $r\to 0$ at a finite viscosity $\muf$,
$ \bra{\delta \ua(r)} \sim r$
since velocity gradients are finite. 
But if we first take the limit $ \nu \to 0$ and then $r\to 0$
($\nu \equiv \muf/\rhof$ is the kinematic viscosity) 
$ \bra{\delta \ua(r)} \sim r^{h}$ with $h\approx 1/3$.
The velocity field is \textit{rough}.
In contrast, ET is, by definition,  a phenomenon at a finite viscosity (small
Reynolds number), thus, the limit $\nu \to 0$ does not make sense
-- the velocity field is always smooth.
But the non-trivial nature of ET manifests itself in the first
subleading term in the expansion~\eq{eq:uexpand}, and this
is best revealed not by  the velocity differences, but by
the second difference of velocity. 
This is the first important result of our work.

\subsection{Intermittency based on velocity differences}
The crucial lesson to learn from the previous section is that
in ET, to uncover the non--trivial scaling of velocity differences  
we must use the second differences of velocity rather than the usual
first difference.
Other than this peculiarity, the rest of this section follows
the standard techniques~\citep{Fri96} used to study 
intermittency/multifractality.

We define the $p$-th order structure function of the second difference of
velocity across a length scale $r$ as: 
\begin{equation}
    \Sigmap(r) \equiv \bra{\lvert\deltat u(\rr)\rvert^p}\/.
    \label{eq:Sigmap}
\end{equation}
We show a representative plot of $\Sigmap$'s for all integer
$p = 1,...,6$ in~\subfig{fig:Scalings}{d} for $\De = 9$. 
Clearly, there exists a scaling regime   
for which the scaling exponents $\zetap$ can be extracted by
fitting $\Sigmap(r) \sim r^{\zetap} $ as $r  \to 0$.
The scaling exponents as a function of $p$ are shown
in~\subfig{fig:Scalings}{e}, where we have
also included half-integer values of $p$.

To obtain reasonable error bars on $\zetap$, we have proceeded in the
following manner:
first we find a suitable scaling regime for each order by visual inspection;
next, in  these chosen ranges, we find the local slopes of the log-log 
plot of $\Sigmap(r)$ vs $r$,
to obtain  $\zetap$ as a function of $r$:
$ \zetap(r) = (\Delta \log \Sigmap (r))/(\Delta \log r)$.
This process is repeated for multiple time snapshots (two successive snapshots are separated by
at least one eddy turnover time) of the velocity field data.
The mean value over the set of exponents thus obtained is the 
exponent $\zetap$ in \subfig{fig:Scalings}{e}  and the standard deviation 
sets the error bar which are shown as a shaded region. 
Clearly, $\zetap$ is a non-linear function of $p$. 
This unambiguously establishes the existence of intermittency in ET.

Furthermore, we have confirmed that the structure functions $\Sp$ of even order
up to $6$ grow as $r^{p}$ for small $r$,
see the Supplementary Material section B3 for discussion.
Thereby we confirm, following the prescription in
Ref.~\citep{schumacher2007asymptotic}, 
that the structure functions of all order are analytic. 
The structure functions begin to depart from this analytic scaling at
a scale that depends very weakly on $\De$ (if at all) but
this scale decreases as $p$ increases. 
We also use another forcing scheme~\citep{esw88} and calculate the
exponents $\zetap$ for $p=2$ and $4$, marked as \RKS{two + symbols in red colour in~\subfig{fig:Scalings}{e}}.
Within errorbars they agree with the values we have
obtained suggesting that the $\zetap$ are universal.

Let us again emphasize that intermittency
is a fundamental property of structure functions, both $\Sp$ and
$\Sigmap$.
The use of $\Sigmap$ is merely to help us extract the exponents $\zetap$.

\subsubsection{PDF of velocity differences}
\begin{figure}
  \centering
  \includegraphics[width=0.85\textwidth]{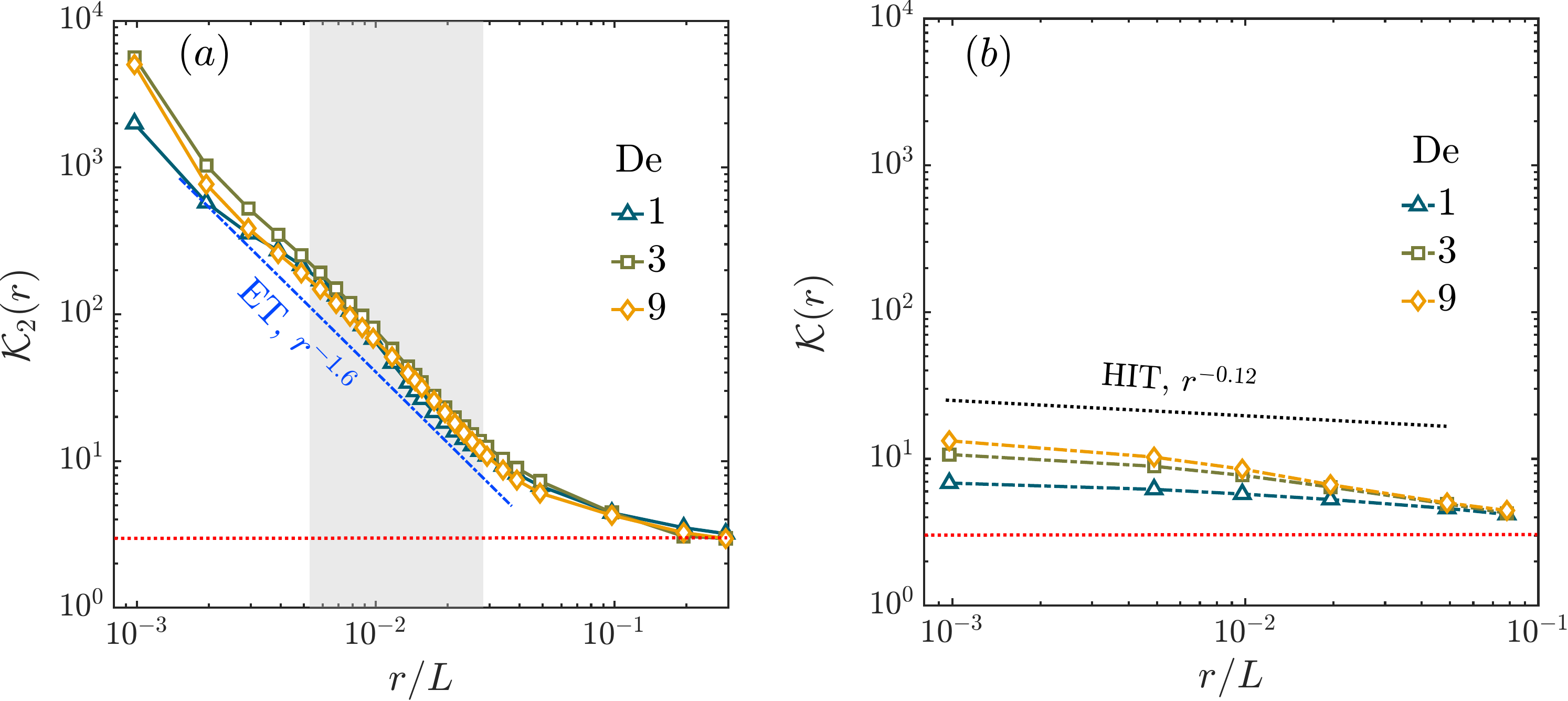}
  \caption{\textbf{Kurtosis} The kurtoses, (a) $\mKt$ and (b) $\mK$ as a
    function of  the scale $r$ for $ \De = 1$, $3$ and $9$.
    The red dashed line is at ordinate equal to $3$. 
    We also show in (a) a line of slope $-1.6$.
    The scaling exponent of kurtosis, obtained from fitting
    the data in the gray shaded region, are:
    $-1.6 \pm 0.3$, $-1.6 \pm 0.1$, and $-1.6 \pm 0.1$, for
    $\De = 1,3$, and $9$, respectively.
    This demonstrates both the non-Gaussian nature of the PDFs and the
    universality of the exponents with respect to $\De$.
    The kurtosis of $\delta u$, $\mK$, grows slower as $r\to 0$
      and may not be universal.
    To compare, we also plot, in (b), corresponding result for
    Newtonian HIT.
    Both the kurtoses $\mK, \mKt \to 3$ (shown in dotted-red line) as $r\to L$.
    This indicates that at large separations the statistics of velocity
    difference are close to a Gaussian.} 
  \label{fig:SK}
\end{figure}
Another way to demonstrate the effects of intermittency is by looking at
the PDF of velocity differences across a length scale.  
From the structure function we have obtained intermittent behavior for scales
$r/L < 1$.
Thus, we expect the PDF of velocity differences to be close to Gaussian for
$r/L \approx 1$ and
to have long tails (decaying slower than Gaussian) for $r/L < 1$. 
This indeed is the case, as is shown in \subfig{fig:Scalings}{f} where we plot
the PDFs of $\deltat u(r) $ for different separations $r$ (for $\De = 9$).
The tails of the distribution of $\deltat u$ decay much slower than
Gaussian, thereby clearly demonstrating intermittency.

Note that also the PDF of the usual velocity differences $\delta u(r)$
is non-Gaussian, see the Supplementary Material Fig.~(S6).
But the PDF of $\deltat u$ falls off slower than
$\delta u$ at the same $r$, in other words larger fluctuations are more likely
to appear in the second difference of velocity -- it is
more intermittent. This non-Gaussianity of probability distributions can be quantified by the Kurtosis (also called
Flatness) defined by 
\begin{equation}
  \mK(r) =\frac{\bra{\left[\delta u(r)\right]^4}}
     {\bra{\left[\delta u(r)\right]^2}^{2}}\/ \quad
  \text{and}\quad
  \mKt(r) = \frac{\bra{\left[\deltat u(r)\right]^4}}
      {\bra{\left[\deltat u(r)\right]^2}^{2}},
\end{equation}
for the first and second difference of velocity, respectively.
For Gaussian distributions the Kurtosis is $3$. 
We find $\mKt \approx 3$ as $r \to L$, i.e., the PDFs (of $\deltat u$)
are close to Gaussian for large separations.
From the scaling behaviour of structure function we obtain
$\mKt(r) \sim r^{\zeta_{\rm 4} - 2\zetat}$ as $r \to 0$,
which is consistent with
$\mKt(r) \sim r^{-1.63}$
obtained from the distribution of second differences shown
in \subfig{fig:SK}{a}.
Furthermore, we find that the Kurtosis is independent of $\De$.
The gray shaded region marks the range used to compute the scaling
  exponents for $\mKt(r)$.
  We obtain: $-1.6 \pm 0.3$, $-1.6 \pm 0.1$, and $-1.6 \pm 0.1$ for
  $\De = 1, 3$ and 9 respectively.
This is further evidence in support of universality of
intermittency in ET. 
The Kurtosis of first difference $\mK(r)$ is also close to a Gaussian
as $r\to L$, but grows much slower than $\mKt(r)$ as $r\to 0$ as shown 
in \subfig{fig:SK}{b}.
In Newtonian HIT at high $\Rey$, the most recent
DNS~\citep{iyer2020scaling,iyer2017reynolds} shows
$\zeta_{\rm 4}-2\zetat \approx -0.12$ so that 
$\mathcal{K} \sim r^{-0.12}$.
Hence, the intermittency we obtain in ET is more intense
than what is observed in HIT.

We also calculate the cumulative PDF (CDF) of $\deltat u$
by rank--order method,
thereby avoiding the usual binning errors that appear
while calculating PDFs via histograms. 
In section C1 of the Supplementary Material we show that
rescaling the abscissa of the CDFs by the
root-mean-square value of $\deltat u$ does not collapse
the CDFs for different $r$, i.e., the PDFs
are not Gaussian. 

Altogether the PDFs of $\deltat u$ provide us three additional evidences
in support of intermittency in ET.

\subsection{Intermittency based on dissipation}
\begin{figure*}
  \centering
  \includegraphics[width=.9\textwidth]{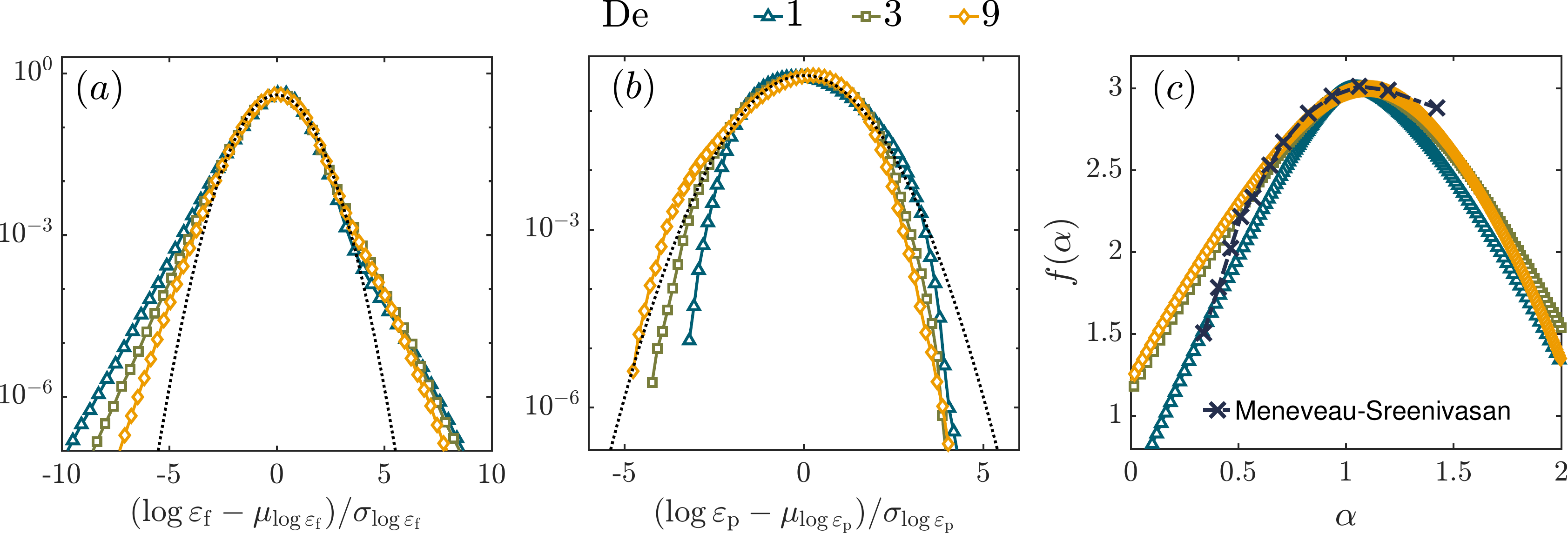}
  \caption{\textbf{Dissipation rates.}
    (a) PDFs of the logarithm of the fluid energy dissipation rate
    $\epsilonf$ for all three $\De$ numbers. We denote by $\mulonfp$ and $\siglonfp$ the mean and variance of the logarithm of energy dissipation rates.
    The distributions deviate significantly from a log-normal behaviour in
    both the left and right tails.
    The right tails coincide for large $\De$, similar to the coincident right
    tails at large $\Rey$ in Newtonian HIT.
    (b) PDFs of the logarithm of the polymer energy dissipation rate
    $\epsilonp$ for all three $\De$ numbers.
    The PDFs of $\log\epsilonp$ are sub-Gaussian, i.e. decay faster than a
    Gaussian, indicating $\epsilonp$ is not intermittent.
    (c) The multifractal spectra of the fluid dissipation field calculated
    from the scaling of the energy dissipation rate $\epsl $ calculated over a
    cube of side $r$.
    The black dash-dotted line shows the spectrum for Newtonian
    HIT~\cite{meneveau1991multifractal}.}
    \label{fig:PDFs}
\end{figure*}
In HIT (high $\Rey$, Newtonian turbulence) there are two routes to
study intermittency:
one is through structure functions and another is through
the fluctuations of the energy dissipation rate~\cite{Fri96} --
the PDF of $\epsilonf$ deviates strongly from a log-normal
behaviour~\citep{Shivamoggi89}.
We now take the second route for ET, in which case there
are two contributions to the total energy dissipation --
$\epsilonf$ and $\epsilonp$. 
In \subfig{fig:PDFs}{a} and \subfig{fig:PDFs}{b} we plot the PDFs of the
logarithm of $\epsilonf$ and $\epsilonp$, respectively.
We find that the former decays slower than a Gaussian,
i.e., the PDF itself falls off slower than a log-normal, whereas
the latter decays faster than a Gaussian.
The fact that the PDF of $\epsilonp$ falls off much faster
than that of $\epsilonf$ can even been seen by
comparing \subfig{fig:snaps}{a} with \subfig{fig:snaps}{b}.
Clearly, the statistics of $\epsilonp$ are
non-intermittent.
Henceforth, following the standard analysis pioneered by
\citet{meneveau1991multifractal} for HIT, 
we study  the scaling of the $q$-th moment of the viscous dissipation averaged
over a cube of side $r$,
\begin{subequations}
  \begin{align}
  \bra{\epsl^q} &\sim r^{\lamq}, \quad\text{where}\\ 
  \epsl &\equiv \frac{2\muf}{\rhof}\bra{\Sab\Sab}_{\rm r}\/.
  \label{eq:epsfl}
  \end{align}
\end{subequations}
Here the symbol $\bra{\cdot}_{\rm r}$ denotes averaging over a cube
of side $r$. 
The Legendre transform of the function $\lamq$ gives the multifractal
spectrum (also called the Cramer's function) $f(\alpha)$:  
\begin{align}
    \lamq = \inf_\alpha \left[ q(\alpha -1) + 3 - f(\alpha) \right],
\end{align}
where singularities in the dissipation field with exponent $\alpha-1$ lie on
sets of dimension $\alpha$.
We plot the $f(\alpha)$ spectrum for ET in \subfig{fig:PDFs}{c}.
There are minor differences between the multifractal spectrum for
$\De = 3$ and $9$ on one hand and $\De = 1$ on the other hand.
The clear collapse of the multifractal spectra at large $\De$ hints towards a
universal multifractality in ET in the limit of large $\De$. 
For comparison, we also plot, in \subfig{fig:PDFs}{c} the multifractal spectrum 
for HIT as a black dash-dotted curve~\cite{meneveau1991multifractal}.
In HIT the intermittency model based on velocity are closely
connected to the intermittency models based on dissipation~\citep{Fri96}.
The development of such a formalism for ET, although important, is
not considered in this work. 

\section{Discussion and summary}

We note that the phenomenon of elastic turbulence has no
Newtonian counterpart -- in the absence of the polymers this 
phenomenon disappears. 
Nevertheless, as HIT is the model of turbulence that has
been studied in great detail we have used it as an illustrative example
to compare with ET.
Such comparison must be done with care. 
In HIT, the theory of Kolmogorov helps us understand the simple scaling
of the energy spectrum, although a systematic derivation starting from the
Navier--Stokes equation is still lacking.
The key insight of Kolmogorov theory is that the energy flux
across scales, due to the nonlinear advective term, is a constant.
In practice, the flux is a fluctuating quantity, where 
its mean value determines the simple scaling prediction $\zetap = p/3$, while
the fluctuations of the flux is the reason behind intermittency.
The fluctuations of the flux shows up as fluctuations of the
energy dissipation rate (because the advective term conserves energy)
which is multifractal.

Elastic turbulence was first discovered at the start of this century.
Almost all studies of ET, so far, have concentrated on understanding the
scaling of the energy spectrum. 
A theory at the level of Kolmogorov theory for HIT is still lacking. 
Nevertheless, it is clear that the mechanism of ET is
very different from HIT.
In the latter, it is the nonlinear advective term that is responsible
for turbulence, while in the former the advective term is expected
to be subdominant, it is the stress from the polymers that
must balance the viscous dissipation in the range of scales where
ET is found.
We show that this is indeed the case.
A consequence of this balance is that the scaling exponents of $E(k)$ and
$\EP(k)$ are related to each other by \eq{eq:XiChi}.
In ET there is not one but two possible mechanisms of energy dissipation.
A particularly intriguing result we obtain is that
only one of them, $\epsilonf$, shows intermittent behaviour, since the
energy dissipation rate due to the polymers is not intermittent, and its logarithm remains sub-Gaussian.

In summary, we have shown that both the velocity field and the
energy dissipation field in ET are intermittent/multifractal.
But this multifractality is very different from the multifractality
seen in HIT.
In HIT, in the limit of viscosity going to zero,
the velocity field is rough. 
In contrast, the velocity field in ET is smooth at leading order, and
roughness and the multifractal
behavior appears due to the sub-leading term.
Consequently, although the velocity difference across a length scale
is intermittent, 
it is necessary to use the second difference of velocity, to
properly reveal the intermittency.
Finally, note that in HIT, the multifractal exponents are expected
to be universal, i.e., they are independent of the method of stirring and the
Reynolds number (in the limit of large Reynolds number).
In ET the multifractality appears at small $\Rey$ and large $\De > 1$.
All the evidences from our DNSs suggest that
intermittency in ET is also universal with respect to Deborah number,
method of stirring and choice of model of polymers,
although significant future work with high resolution DNSs
are necessary to provide
conclusive evidence.

\subsection*{Methods}

We solve eqns.~\ref{NN},~\ref{Conf} using a second order central-difference 
scheme on a $L=2\pi$ tri-periodic box discretized by $N^3 = 1024^3$
collocation points, such that $L/N = \Delta \approx 0.05 \eta$, 
where $\eta\equiv (\nu^3/\bra{\epsilonf})^{1/4}$ is 
the Kolmogorov dissipation length scale and 
$\nu\equiv \muf/\rho$ is the kinematic viscosity.
Integration in time is performed using the second order Adams-Bashforth
scheme with a time step $\Delta t \approx 10^{-5} \tauK$, 
with $\tauK\equiv \sqrt{\nu/\bra{\epsilonf}}$.
We use $18$ snapshots in our analysis, with successive snapshots separated by 
$\approx 1.6 \times 10^4\tauK$. We choose a $\muf$ so as to obtain a laminar 
flow in the Newtonian case with $A=B=C=1$
-- this corresponds to the Taylor scale Reynolds number
  $\Rey_\lambda \approx 40$.
Next, we choose $\mup$ such that the viscosity ratio
$\muf/(\muf+\mup)=0.9$ -- this corresponds to  dilute
polymer solutions~\cite{Prasad10},
and we vary $\taup$ over two order of magnitudes.
The numerical solver is implemented on the in-house code \textit{Fujin}; see \url{https://groups.oist.jp/cffu/code} for additional details and validation tests. The very same code has been successfully used on various problems involving Newtonian and non-Newtonian fluids \citep{abdelgawad2023scaling, soligo2023non, rosti2023large, aswathy_rosti_2024a}. 

\subsection*{Data availability}
All data needed to evaluate the conclusions are present in the paper and/or the Supplementary Materials, and available as a Source Data file at \url{https://groups.oist.jp/cffu/singh2024natcommun}. 

\subsection*{Code availability}
The code used for the present research is a standard direct numerical simulation solver for the Navier--Stokes equations. Full details of the code used for the numerical simulations are provided in the Methods section and references therein.

\bibliography{references,turb_ref}

\subsection*{Acknowledgments}
The research was supported by the Okinawa Institute of Science and Technology
Graduate University (OIST) with subsidy funding from the Cabinet Office,
Government of Japan. The authors acknowledge the computer time provided by
the Scientific Computing \& Data Analysis section of the Core Facilities at OIST and
the computational resources provided by the HPCI System (Project IDs: hp210229, hp210269, and hp220099).
The authors RKS and MER thank Prof. Guido Boffetta for crucial insights and
suggestions and for bringing to our notice Ref.~\citep{Biferale01}. PP 
acknowledges support from  the Department of Atomic Energy (DAE), India under
Project Identification No. RTI 4007, and DST (India)
Project No. MTR/2022/000867.
DM acknowledges the support of the Swedish Research Council Grant No.
638-2013-9243. NORDITA is partially supported by NordForsk.
DM gratefully acknowledges hospitality of OIST. 

\subsection*{Author contributions}
M.E.R. and D.M. conceived the original idea. M.E.R. planned and supervised
the research, and developed the code. R.K.S. and M.E.R. performed the
numerical simulations. R.K.S. analyzed the data.
R.K.S. and D.M. wrote the first draft of the manuscript.
All authors outlined the manuscript content and wrote the manuscript.

\subsection*{Competing interests}
The authors declare that they have no competing interests.

\appendix


\pagenumbering{gobble}

\setcounter{table}{0}
\makeatletter 
\renewcommand{\thetable}{S\@arabic\c@table}
\makeatother

\setcounter{figure}{0}
\makeatletter 
\renewcommand{\thefigure}{S\@arabic\c@figure}
\makeatother

\setcounter{equation}{0}
\makeatletter 
\renewcommand{\theequation}{S\@arabic\c@equation}
\makeatother

\section{Energy Spectra and flux}
\label{smat:energy}

We show several crucial aspects of our DNS.

\begin{figure}[!ht]
  \centering
  \includegraphics[width=0.9\textwidth]{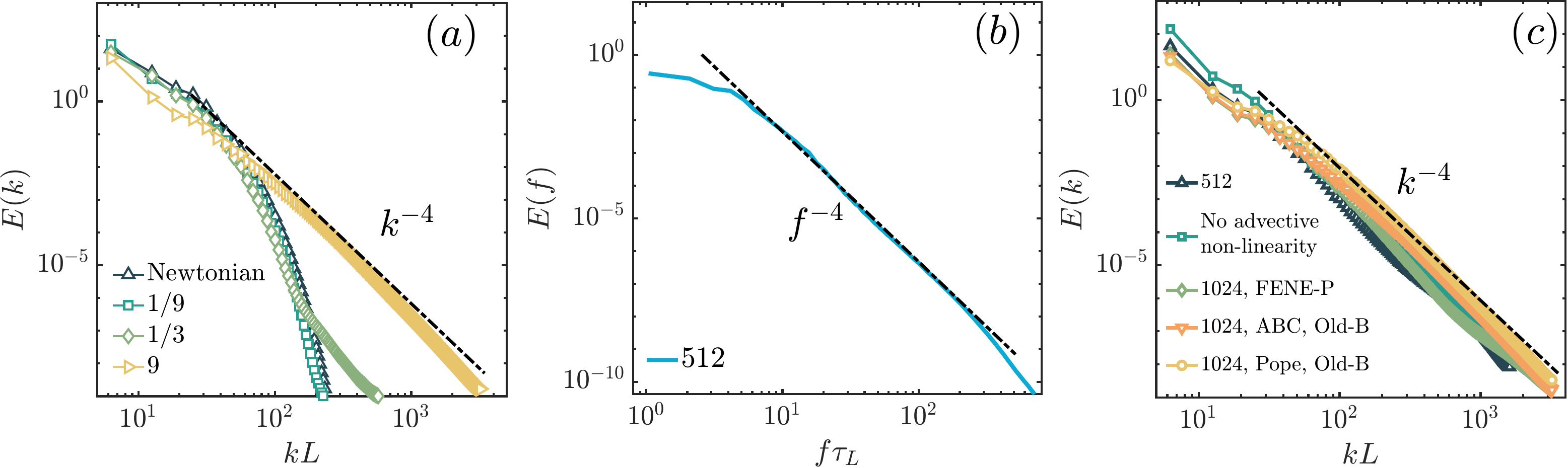}
  \caption{\textbf{Energy spectra.}
      (a) Energy spectra for Newtonian and small $\De$ falls sharply --
    do not show any scaling behavior. This is compared against that for
    $\De = 9$, where the spectrum decays as $k^{-4}$ (see also Fig.3 of main
    text).
    (b) The temporal energy spectrum shows a similar $E(f) \sim f^{-4}$
    power--law behaviour (computed for a smaller grid size $N^3 = 512^3$).
    (c) Universality with respect to grid size, polymer model, and the forcing
    scheme. The same  $k^{-4}$ scaling is obtained for simulations with
    smaller grid size $N^3 = 512^3$, no advective non-linear term,
    FENE-P model of polymers, a different forcing scheme~\citep{esw88}
    --the force is $\delta$-correlated in space and
    exponentially correlated in time.
    For reference, we also show the $k^{-4}$ spectrum for our
    $N^3 = 1024^3$ simulation with the non-linear term using the Oldroyd-B
    model of polymers excited by the ABC forcing scheme (see main text).} 
  \label{fig:Ef}
\end{figure}

\begin{enumerate}
\item \textbf{Elastic turbulence appears at $\De > 1$}\\
We show the energy spectrum $E(k)$ for the Newtonian and small
$\De$ (=1/9, 1/3) flows in \subfig{fig:Ef}{a}, which remain devoid of any
appreciable scaling regime.
Energy is concentrated in the largest scales and
the energy per mode decays sharply as we go to small scales.

\item \textbf{Energy spectra in wavenumber and frequency domain
  shows same scaling exponent.}\\
In~\subfig{fig:Ef}{b}, we plot the energy spectra, $E(f)$, of the time-series
of velocity measured at a fixed Eulerian point, for $\De = 1$.
A clear scaling regime of $E(f) \sim f^{-4}$ spans more than a decade in
frequencies $f$. 
This energy spectrum is obtained by applying a
Hanning window to the velocity
field time-series recorded at a single point, which is then averaged 
over numerous ($128^2$) such points in the flow domain. 

\item \textbf{Results are independent of grid resolution.}
We also show the independence of our results from the choice of grid
resolution in~\subfig{fig:Ef}{c}, where the plots of fluid energy spectra for
$\De =1$ from simulations of grid sizes $N = 512$ (blue) and $N = 1024$
(yellow) closely follow each-other.

\item \textbf{The advective nonlinearities are not responsible for ET}\\
We show the spectra from a simulation
where the nonlinear term was set to zero. 
This confirms that ET we observe is indeed sustained by purely
elastic effects.

\item \textbf{Universality with respect to force.}\\
We find that the spectral exponent 
remains unchanged for two different forcing scheme (random forcing and ABC). 
We  also find the scaling exponents for $\Sdiff$ and
$\Sifour$ to be $2.94 \pm 0.21$ and $4.32 \pm 0.27$ respectively when a 
random forcing is used, which are well within the error bars of the 
exponents computed when the forcing was ABC.

\item \textbf{Universality with respect to model for polymer.}
We also obtain the same exponent for energy spectra for 
the  Oldroyd-B model, which we use in all the simulations
in our paper, and the FENE-P model.
\end{enumerate}
\subsection{Fluxes}
\label{smat:fluxes}
\begin{figure}[!ht]
  \centering
  \includegraphics[width=0.5\textwidth]{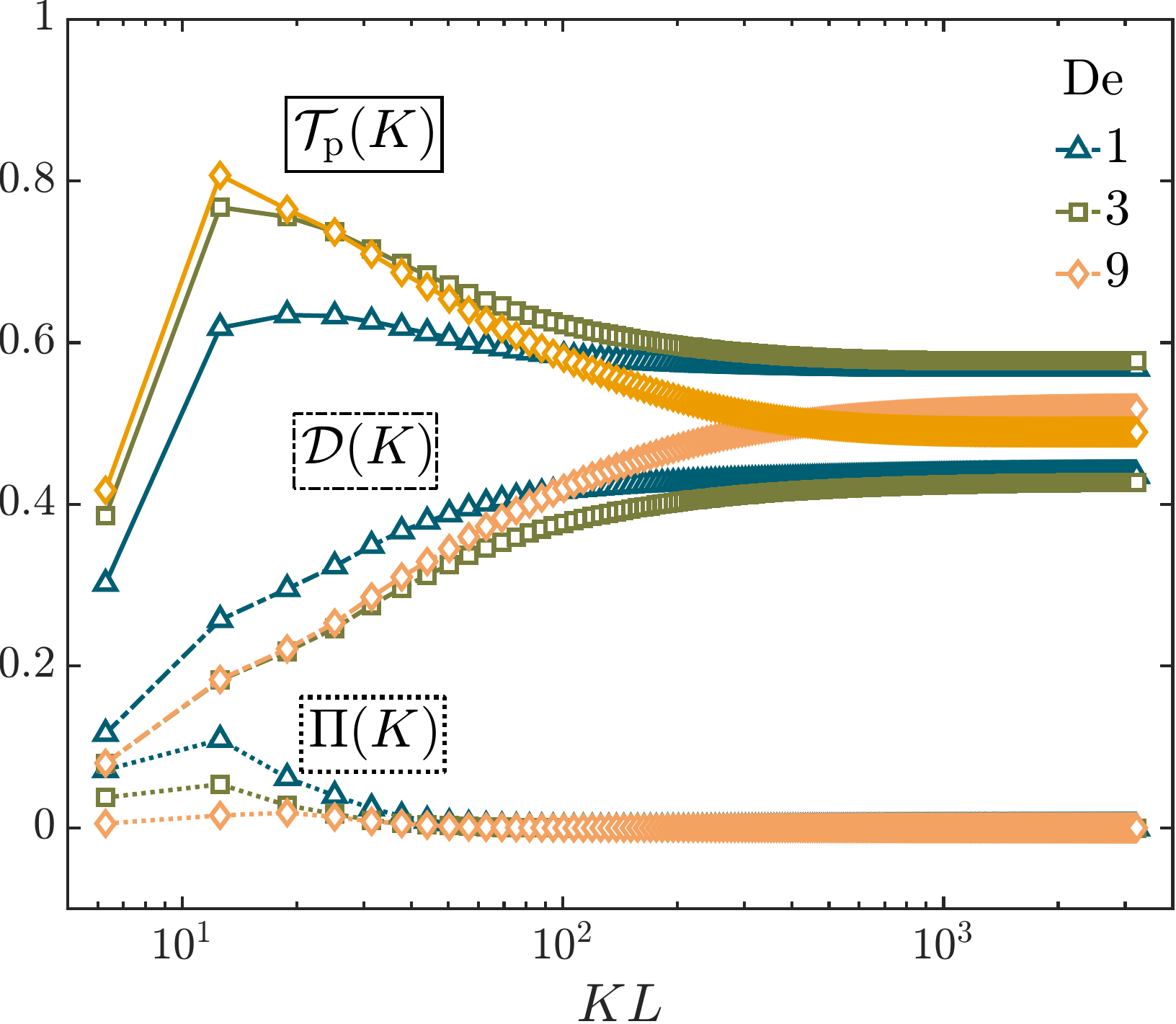}
  \caption{\textbf{Energy fluxes.} The polymeric contribution
    $\Tp(K)$ to the energy transfer to smaller scales compared
    to the fluid dissipation $\mD(K)$ for all three $\De$ numbers.
    Note the near zero contribution of the nonlinear advective  flux
    for $KL \geq 30$.} 
  \label{fig:Fluxes}
\end{figure}
The contribution to the flux of kinetic energy from 
polymer stress, viscous stress, and advective nonlinearity
are, respectively: 
\begin{subequations}
\begin{align}
	\Tp(K) &= \frac{\mup}{\rhof \taup}\int^{K}_{0} d\Omega \ k^2 \ dk \ \ua(\kk) k_{\beta} \Cab(-\kk)\/, \\
	\Tnu(K) &=  \frac{\muf}{\rhof}\int^{K}_{0} d\Omega \ k^2 \ dk \ k^2 \ua(\kk)\ua(-\kk)\/,\\
    \Pi(K) &= i \int^{K}_{0} d\Omega \ k^2 \ dk \ \ua(-\kk)\   \int d{\bm q} \ q_{\beta} u_\beta({\bm q}) \ua(\kk-{\bm q}).
\end{align}
\end{subequations}
We plot them in~\Fig{fig:Fluxes}.
The fluid dissipation and polymer contributions are, as expected, comparable
over a wide range of scales
while the contribution from the advective nonlinear is almost zero
(for $KL \geq 30$).
This justifies the dominant balance used in the main text. 

\subsection{Scaling argument}
Let us first state a well--known result~\citep[see, e.g.,][section 4.5]{Fri96}
for the spectrum of a statistically stationary and isotropic
random vector function $\ww(\xx)$.
We define
\begin{equation}
  \Gamma(r) \equiv \bra{w_{\alpha}(\xx)w_{\alpha}(\xx+\rr)},
  \label{eq:G1}
\end{equation}
where $r = \lvert \rr \rvert$.
Here we have assumed that the statistics of $\ww$ are isotropic,
hence $\Gamma(r)$ is a function of $r$ alone, not $\rr$.
In this case, the  Wiener--Khinchin formula for the
spectrum, $G(k)$, of $\ww$ is:
\begin{equation}
  G(k) = \frac{1}{\pi}\int_0^{\infty} kr\Gamma(r)\sin(kr) dr.
  \label{eq:Gk1}
\end{equation}
If, in addition, $\ww$ is scale invariant with a scaling exponent $h$,
then under scaling $x \to \lambda x$ we obtain
\begin{equation}
  \ww\to \lambda^h\ww\/,\quad\Gamma\to\lambda^{2h} \Gamma\/,\quad
  k\to\frac{1}{\lambda}k\/.
  \label{eq:scaling}
  \end{equation}
As $\ww$ is scale invariant, its spectrum must be a power-law, hence
we have
\begin{equation}
  G(k) \sim k^{-n}.
  \label{eq:Gk2}
\end{equation}
Substituting \eq{eq:scaling} and \eq{eq:Gk2} to \eq{eq:Gk1} we obtain
\begin{equation}
  2h+1 = n\/.
\label{eq:bridge}
\end{equation}
Strictly speaking, this result holds only for $ 1 < n < 3$.
For $n$ outside this range, \eq{eq:bridge} still holds
under the following conditions:
\begin{enumerate}
\item  $G(k)$ shows power-law scaling with a range of
Fourier modes  $ \knot < k < \Lambda $
where  $\knot$ and $\Lambda$ are the infra-red cutoff and ultra-violet
cutoff, respectively.
\item Outside these cutoffs, $G(k)$
goes to zero fast enough such that
$\int_0^{\infty} dk G(k)$ is finite.
\end{enumerate}
Typically, these conditions are satisfied by
all hydrodynamic quantities. 
Thus, there is a range of scale $ (1/\Lambda) > r > (1/\knot) $
over which $\ww(r)$ is scale-invariant with a scaling exponent $h$
that satisfies~\eq{eq:bridge}.
Also note that the second order structure function of $\ww$
\begin{equation}
  \Stwo(r) = \left[\left\{\ww(\xx+\rr) - \ww(\xx)\right\}
               \cdot\left(\frac{\rr}{r}\right)\right]^2 
  = \frac{2}{3}\left[ \Gamma(0) - \Gamma(r) \right]
  \propto \Gamma(r)
  \sim r^{2h} \sim r^{n-1}.
\end{equation}
 
Let us now consider elastic turbulence where the velocity field
is scale invariant with a scaling exponent $h$ and the tensor $\Cab$
is scale invariant with a scaling exponent $2b$.
This implies that the tensor $\mB$ is scale invariant with scaling
exponent $b$. 
In other words, under rescaling $x\to \lambda x$,
\begin{equation}
  \uu \to \lambda^h \uu \quad\text{and}\quad B_{\ab} \to \lambda^b B_{\ab}.
  \label{eq:ETscaling}
\end{equation}
Let us now assume
\begin{equation}
 E(k) \sim k^{-\Xitwo}\quad \text{and}\quad \EP(k) \sim k^{-\Chit}.
\end{equation}
Applying \eq{eq:bridge} we obtain
\begin{equation}
  2h+1 = \Xitwo \quad\text{and}\quad 2b+1 = \Chit\/. 
\label{eq:xi_chi}
\end{equation}
Note that \eq{eq:bridge} can be extended to apply to the second rank
tensor $\mB$ in a straightforward manner. 
In ET, the advective term in the momentum equation is small (because
$\Rey$ is small) and for small scales the external force in zero.
Hence at small scales, statistical stationarity implies that
we expect the dominant balance to be 
\begin{equation}
  2\muf\Sab  \sim \frac{\mup}{\taup} \Cab
  \label{eq:balance}
\end{equation}
where $\Sab = \frac{1}{2}(\dela\ua + \delb \ub)$.
Applying scale invariance, \eq{eq:ETscaling} to \eq{eq:balance}, we obtain
\begin{equation}
  h-1 = 2b.
\label{eq:hb}
\end{equation}
Finally, substituting \eq{eq:xi_chi} in \eq{eq:hb} we obtain
\begin{equation}
  \Xitwo = 2\Chit + 1.
\end{equation}

\section{Structure Functions}
\label{smat:sf}
The usual structure functions defined by the moments of first differences of
velocity are:
\begin{subequations}
\begin{align}
    \Sp(r) &\equiv \langle \lvert \deltau(\rr) \rvert^p \rangle,   \quad\text{where}\/\\
    \deltau({\bm r}) &\equiv \left[\ua(\xx+\rr) - \ua(\xx)\right]\frac{r_{\alpha}}{\abs{\rr}}\/.
    \label{eq:Sp}
\end{align}
\end{subequations}
From the main text we repeat the definition of the structure function
of second differences
\begin{subequations}
  \begin{align}
    \Sigmap(r) &\equiv \bra{\lvert\deltat u(\rr)\rvert^p}, \quad\text{where}\/\\
    \deltat u({\bm r}) &\equiv \left[\ua(\xx+\rr) -2\ua(\xx) + 
	  \ua(\xx-\rr)\right]\left(\frac{r_{\alpha}}{\abs{\rr}}\right)\/.
    \label{eq:Sigp}
    \end{align}
\end{subequations}

\subsection{Second order structure functions}
\begin{figure}[!ht]
  \centering
  \includegraphics[width=0.95\textwidth]{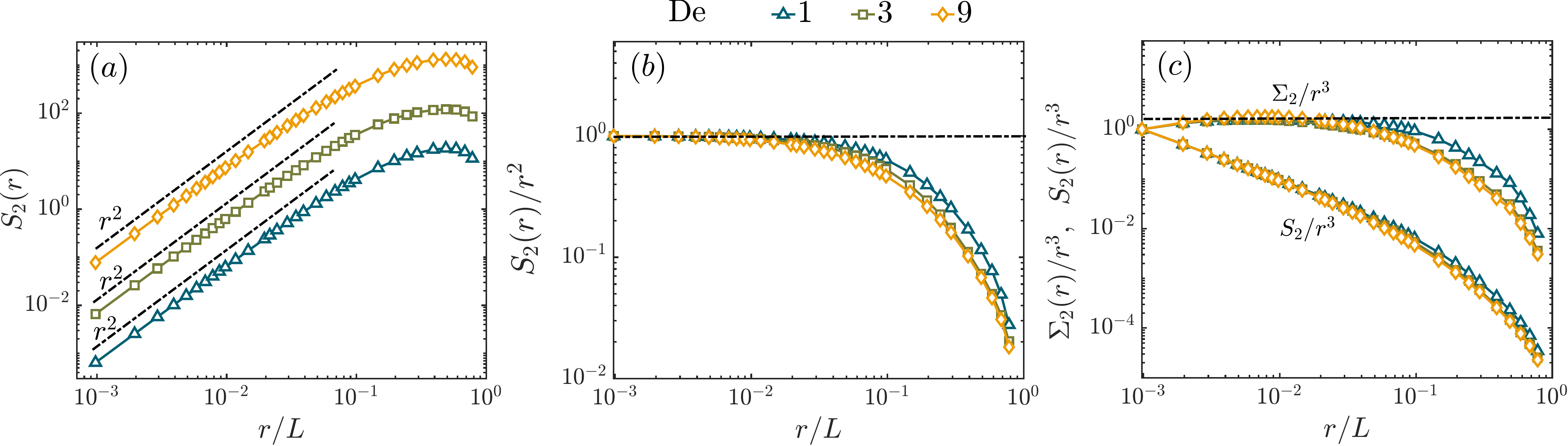}
  \caption{\textbf{Second order structure functions}
    (a) The second order structure function $\Stwo(r)$
    as a function of $r$, for $ \De = 1, 3$, and $9$.
    (b)  $\Stwo(r)/r^2$ for three different $\De$ ranging from $1$ to $9$. The range of analytic scaling $S_p \sim r^p$ decreases with increasing order of the moments as intermittent effects become more important.
    (c) $\Sdiff(r)/r^3$ and $\Stwo(r)/r^3$ as a function of $r$. 
    The range of analytic scaling (shown for $S_4$) remains independent of $\De$.} 
  \label{fig:Stwo}
\end{figure}
In \subfig{fig:Stwo}{a} we show the structure functions $\Stwo(r)$
for different $\De$ as a function of $r$ on a log-log scale.
We obtain the trivial scaling $\Stwo \sim r^2$ as $r\to 0$.
In \subfig{fig:Stwo}{b} we plot $\Stwo(r)/r^2$ for three different
$\De$ ranging from $1$ to $9$.
At small enough $r$ they all show $\Stwo \sim r^2$.
As $r$ increases they all depart from this trivial scaling
at a length scale $\ell$ which depends very weakly on $\De$,
if at all. 
Does the departure from trivial scaling shows a new scaling
range?
From \subfig{fig:Stwo}{a} it is unclear if there is a scaling
range at intermediate $r$.
Now we turn to second order structure function of $\deltat u$, $\Sdiff$. 
We find that $\Sdiff(r)$ shows a significant scaling range
as $r\to 0$ with the non-trivial scaling exponent
$\zetat \approx 3$, see the main text.
To substantiate this further we plot in \subfig{fig:Stwo}{c}
$\Sdiff/r^3$ and $\Stwo/r^3$.
The former shows a plateau confirming $\Sdiff \sim r^3$,
while the latter shows practically no plateau. 
There is no range of scales where the scaling
$\Stwo \sim r^3$ is obtained.

\subsection{Correlation function}
\begin{figure}
  \includegraphics[width=0.4\linewidth]{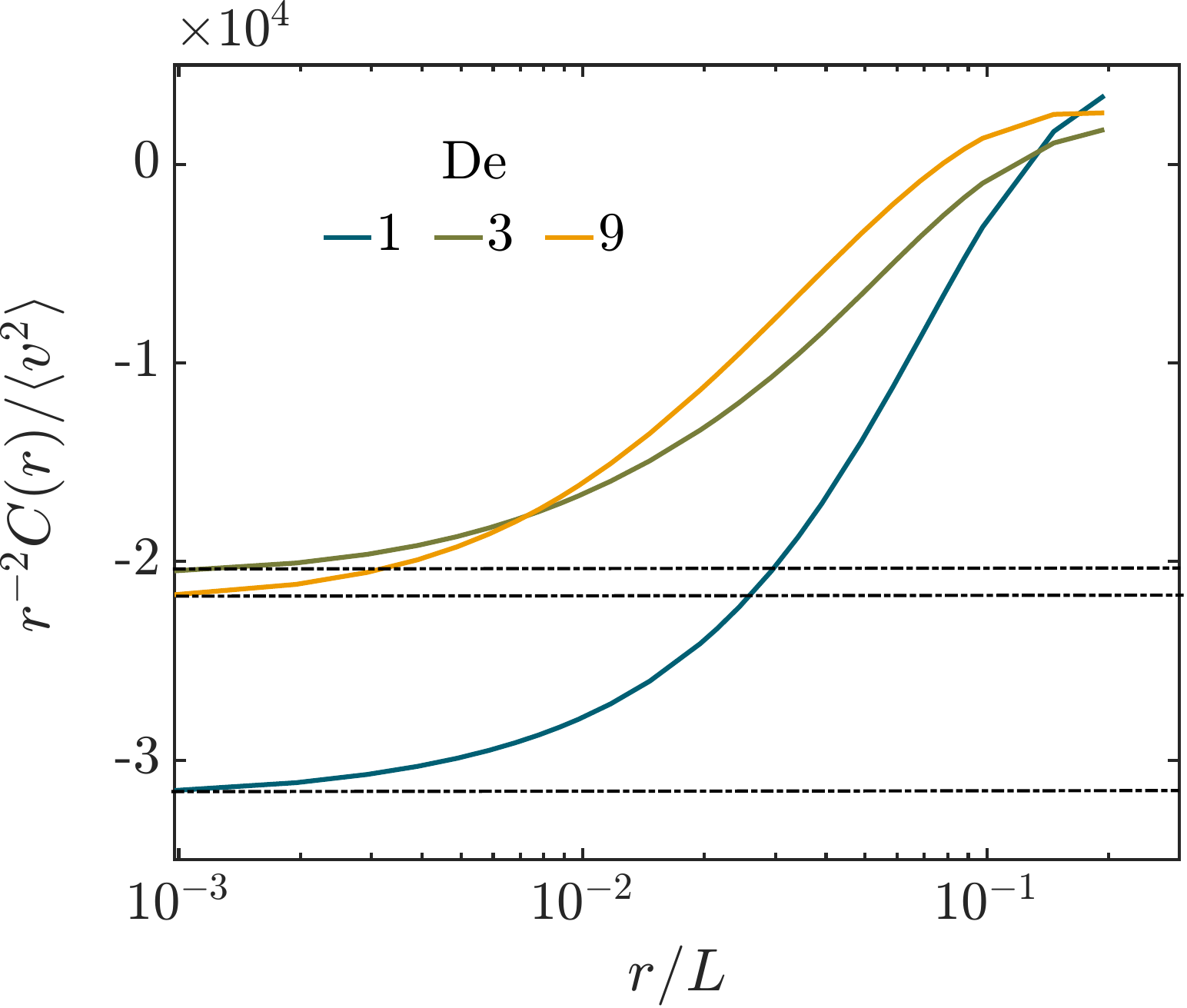}
  \caption{\textbf{Correlation of velocity fluctuations} The semilog plot of the normalized, compensated correlation
    $C(r) \equiv \lrp{\Sdiff(r) - 2\Stwo(r)}/2 = \bra{\delta u(r)\delta u(-r)}$.
    This correlator gets its dominant contribution from $\Stwo(r)$ in the
    limit $r \to 0$, thus recovering the analytic behaviour.}
  \label{fig:C}
\end{figure}
Note that two--point correlation function of velocity
$C(r) \equiv  \bra{\delta u(r)\delta u(-r)}$
is related to $\Sdiff$  by
\begin{equation}
\Sdiff(r) = \bra{[ \delta u(r)+\delta u(-r) ]^2} =
2 \Stwo(r) + 2 C(r).
\label{eq:S2C}
\end{equation}
 Then the consequence of  our results is that:
\begin{equation}
 2 C(r) = \Sdiff(r) - 2\Stwo(r) \sim A r^3 - B r^2,
  \label{eq:C}
\end{equation}
where $A$ and $B$ are two constants. 
Hence in the limit $r\to 0$, $C(r) \sim r^2$.
We check this explicitly by plotting  $C(r)/r^2$ as a function of $r$
on a log-lin scale in \Fig{fig:C} for three $\De$.  
At small $r$, the plot becomes flat, confirming our expectations.

Another way to see this is that, 
we expect the velocity gradients 
to be smooth for small $r$, whereby 
$\ua(\xx+\rr)-\ua(\xx) \equiv \delta \ua(r) \sim r_{\alpha}G_{\ab} $ and 
$\delta \ua(-\rr) \sim -r_{\alpha} G_{\ab}$,
where $G_{\ab}$ is the gradient of the velocity field evaluated at $\xx$. 
Consequently, 
$C(r) \sim -r^2$ and $\Stwo(r)\sim r^2$ have the same leading order behavior, 
and neither reveals the subdominant non-trivial scaling.
\subsection{Higher order structure functions}
\begin{figure}[!ht]
  \centering
  \includegraphics[width=0.95\textwidth]{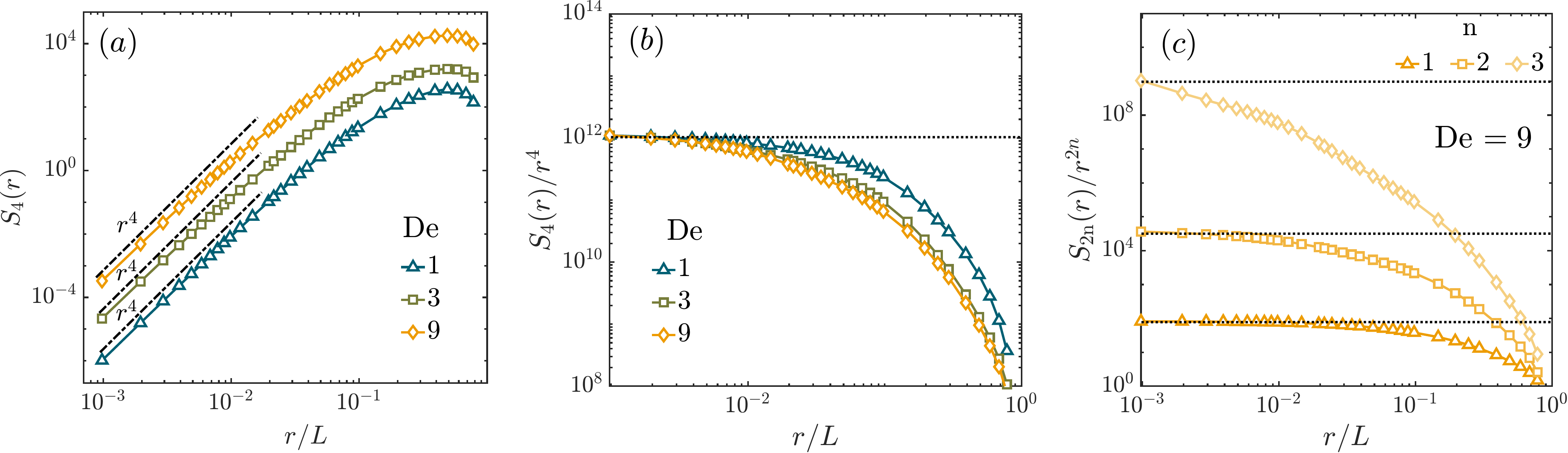}
  \caption{\textbf{Higher order structure functions}
    (a) The fourth structure function $\Sfour(r)$
    as a function of $r$, for $ \De = 1, 3$, and $9$.
    We obtain the trivial scaling $\Sfour(r) \sim r^4$
    as $r\to 0$. 
    (b)  $\Sfour(r)/r^4$ for three different $\De$ ranging from $1$ to $9$.
    (c) $S_{\rm 2n}/r^{2n}$ as a function of $r$ for $\De = 9$
    for $n=1, 2$ and $3$. The range over which the
    trivial scaling is valid decreases as we consider structure
    functions of higher order.} 
    \label{fig:SFs}
\end{figure}
In \subfig{fig:SFs}{a} we plot the fourth structure function $\Sfour(r)$
as a function of $r$, for $ \De = 1, 3$, and $9$.
We obtain the trivial scaling $\Sfour(r) \sim r^4$
as $r\to 0$. 
In \subfig{fig:SFs}{b} we plot  $\Sfour(r)/r^4$ for three different
$\De$ ranging from $1$ to $9$.
We find that all of them depart from trivial scaling at large $r$,
but the scale at this this departure appears is almost
independent of $\De$. 
In \subfig{fig:SFs}{c} we plot the even order structure
functions $S_{\rm 2n}/r^{2n}$ as a function of $r$.
As  $r\to 0$, $S_{\rm 2n}(r) \to r^{2n}$.  
Thereby we confirm, following the prescription in
Ref.~\citep{schumacher2007asymptotic}, 
that the structure functions of all order are analytic. 
The structure functions begin to depart from this analytic scaling at
a scale that depends very weakly on $\De$ (if at all), but
this scale decreases as $p$ increases.
The same behaviour was observed in Ref.~\citep{schumacher2007asymptotic}
for the case of Newtonian HIT. 
\section{Probability distribution function of velocity differences}
\begin{figure}
  \includegraphics[width=0.95\linewidth]{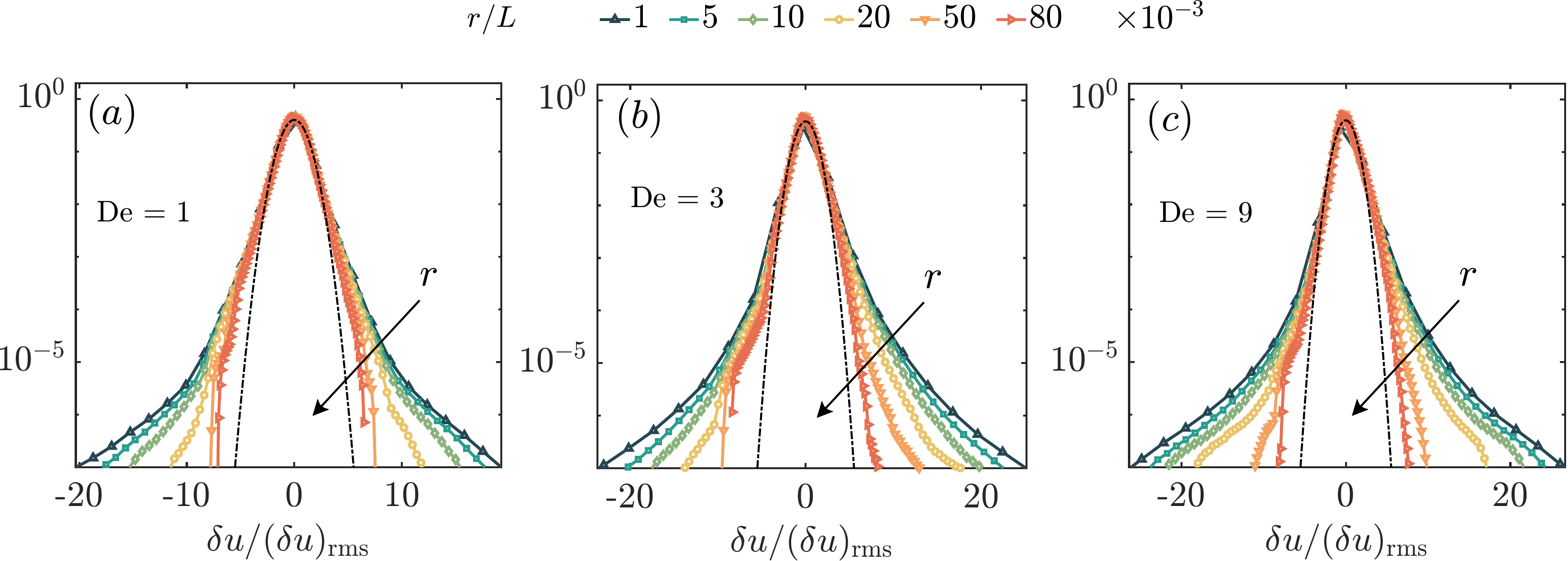}
  \caption{\textbf{Probability distributions of $\bm{\delta u}$} The PDF of $\delta u$ (normalized by
    their root--mean--square value) for different values of $r$,
    for: (a) $\De = 1$, (b) $\De = 3$, and (c) $\De = 9$. For
    comparison, we have plotted an Gaussian distribution as a dashed
  black line.}
  \label{fig:PDF_du}
\end{figure}
\begin{figure}
  \includegraphics[width=0.45\linewidth]{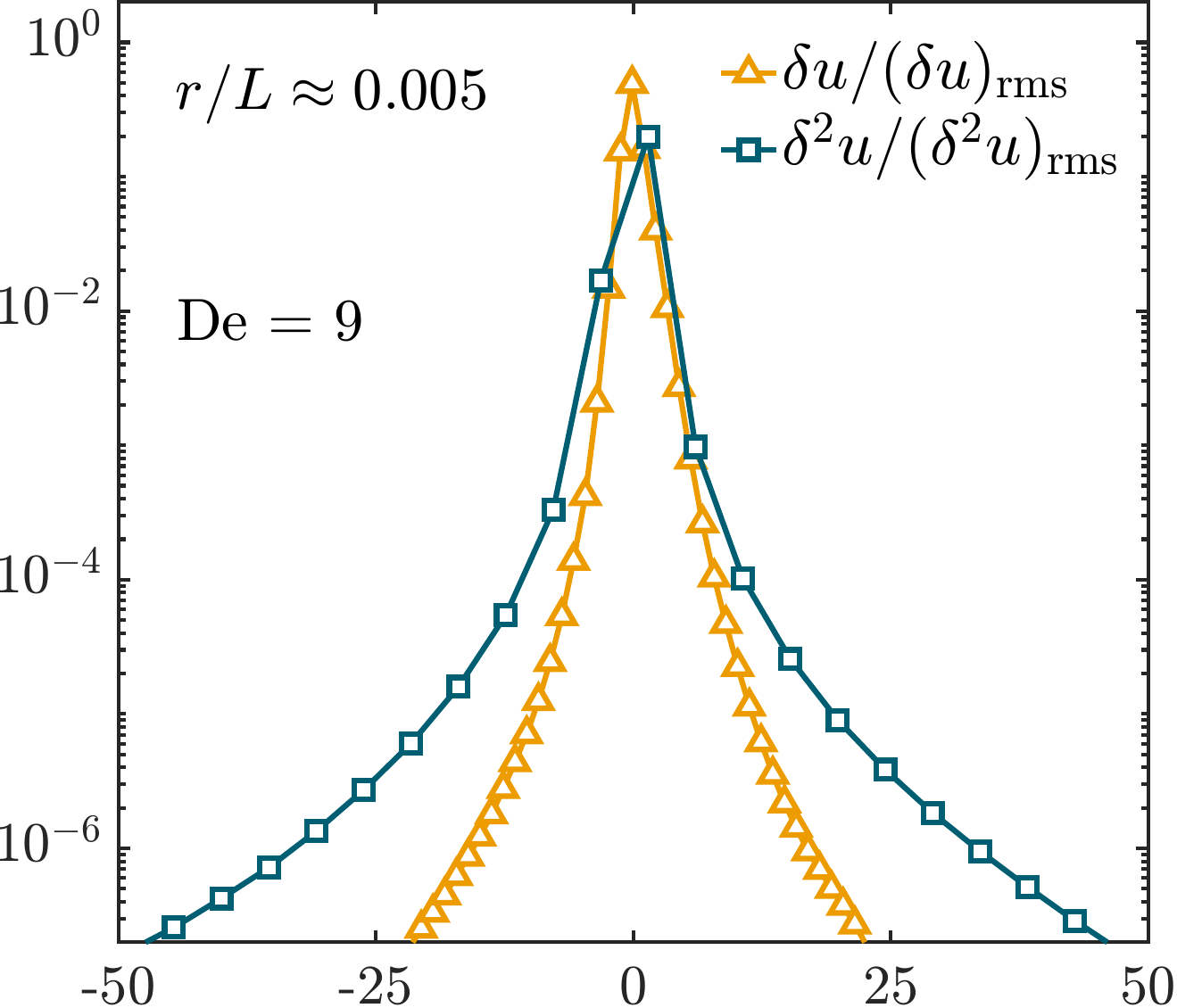}
  \caption{\textbf{Intermittency in velocity differences} The PDF of $\deltat u(r)$ and $\delta u(r)$ for a representative
    value of $r$ in the scaling range. The tail of the PDF of second
    difference falls off much slower than that of the first difference.
    This clearly indicates that the second differences are more intermittent
    than the first.}
    \label{fig:comp_PDF_du}
\end{figure}

An evidence of intermittency is non-Gaussian behaviour of the
tail of the PDF of velocity differences across a length scale.
In the main text we show the PDF of second difference of velocity,
$\deltat u$ across a length scale $r$.
These PDFs are non-Gaussian if the scale $r$ falls within the
scaling range, $r/L \gg 1$, of the structure functions.
In \Fig{fig:PDF_du} we plot PDF of $\delta u$ for several different
values of $r$ for the three values of $\De$.
These PDFs are non-Gaussian too.
But they are less intermittent than the corresponding PDF of
the second difference of velocity $\deltat u$,
see \Fig{fig:comp_PDF_du} where we plot the two PDFs for
a fixed representative value of $r$ within the scaling range. 
This demonstrates what we have already commented on, the intermittency
is a fundamental property of the velocity difference,
but it is best revealed by the second difference, $\deltat u$.
 
\subsection{Cumulative probability distribution}
\label{smat:cdf}
\begin{figure}
  \centering
  \includegraphics[width=0.95\textwidth]{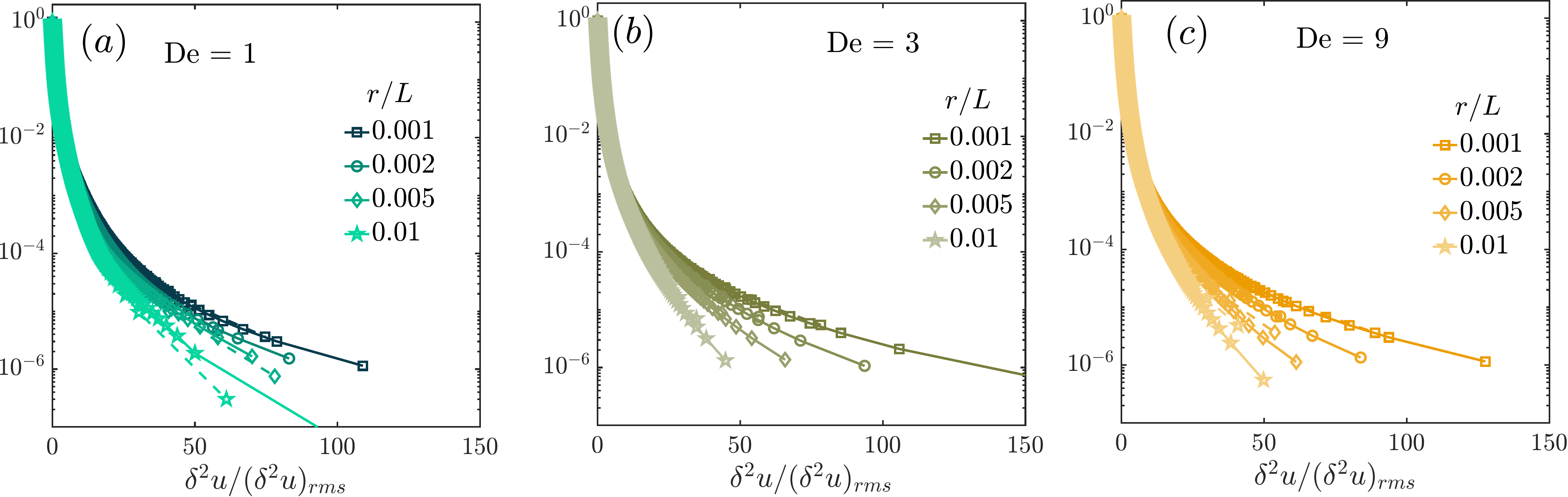}
  \caption{\textbf{Complementary Cumulative Distribution} CCD functions of
    $x$ component of $\deltat u$ (normalized
   by their root-mean-square value) for (a) $ \De = 1$, (b) $3$, and (c) $9$.
   The solid and dashed curves correspond respectively to CCDs for
   $F^+$ and $F^{-}$, calculated by rank-order method.}
  \label{fig:CDFs}
\end{figure}
We now consider two complementary cumulative probability distribution
  (CCD) 
functions 
\begin{equation}
	F^{+}(X) = 1-\int_0^{X} P(x)dx\quad\text{and}\quad 
	F^{-}(X) = 1-\int_{-X}^{0} P(x)dx,
\end{equation}
where $X$ is positive and $P(x)$ is the probability distribution function of 
$x$.
The two CCDs characterise the positive and negative tails of the PDF
respectively. 
In ~\Fig{fig:CDFs} we plot them, $F^{\pm}(\deltat u)$, for the second
differences of velocity.
The solid lines correspond to the CDFs computed for $\deltat u \geq 0$,
while the dashed curves correspond
to CDFs for $\deltat u \leq 0$.
The CDFs are calculated using rank-order method~\citep{mitra2005multiscaling},
thereby they are free of binning errors that
plague the usual PDFs that are calculated from histogram. 

\section{Computing Exponents}
\label{smat:local}

\begin{figure}
  \centering
  \includegraphics[width=0.95\textwidth]{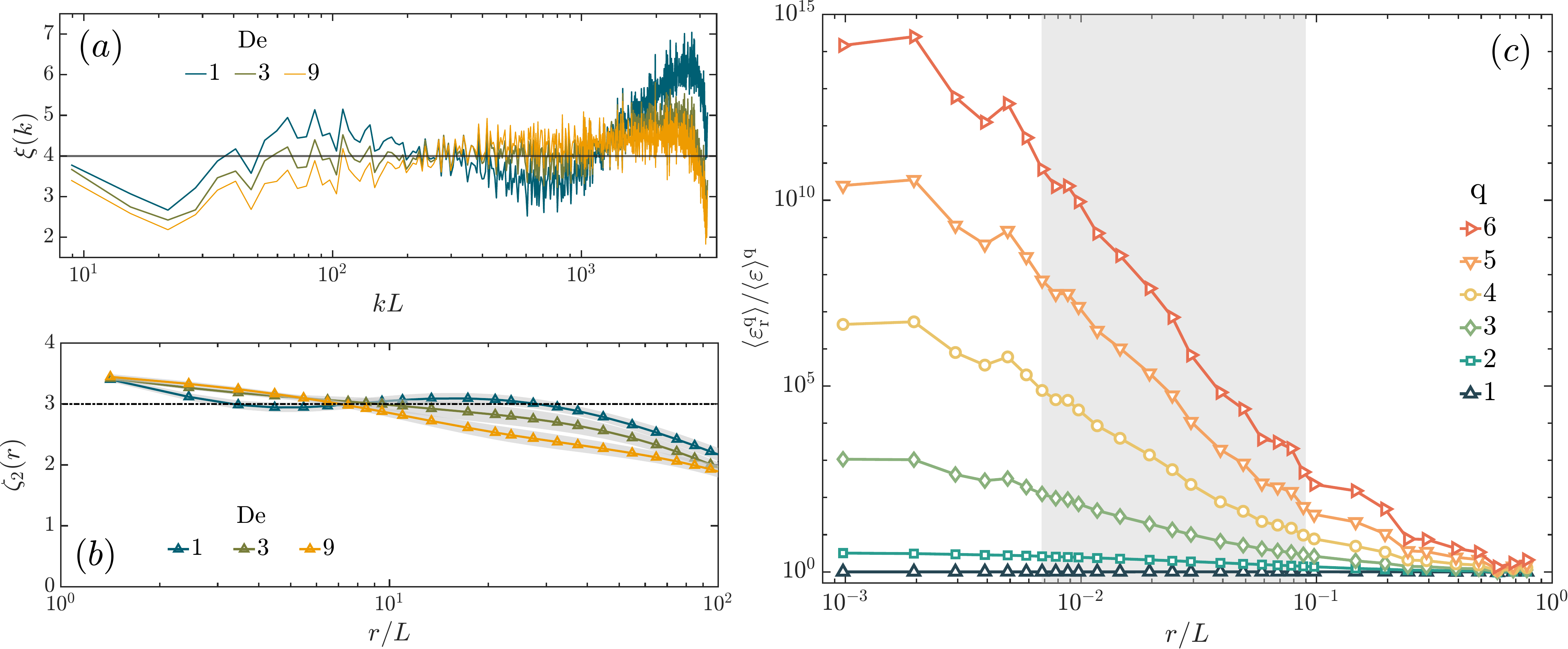}
  \caption{\textbf{Exponents of the power-law scalings} (a) The local slopes of energy spectra for $\De =  1, 3,$ and $9$ after a 3-point moving average. (b) The local slopes $\zetat$ for $\Sdiff(r)$ for all three $\De$. The solid curves correspond to the mean value of the exponents with the standard deviation shown as shaded region. The statistics were obtained using 18 field snapshots.  (c) The log-log plot of the scale averaged fluid energy dissipation rate $\bra{\epsl^q}$ versus the scale ${\rm r}$ for $\De = 3$.}
  \label{fig:Loc_slop}
\end{figure}

In this section, we detail the computation of scaling exponents of structure functions and energy spectra in terms of local slopes of the log-log curves. These have been shown in panels (a) and (b) of \Fig{fig:Loc_slop}. We show in panel (c) the scale-averaged energy dissipation rate $\bra{\varepsilon_r}$.

The scaling exponents $\xi$ of the energy spectrum $E(k)$ shown in panel (a) are the 3-point moving averages of their local slopes. The mean exponents and their standard deviations are then found to be $-4.0 \pm 0.6, -4.0 \pm 0.3$, and $-4.0 \pm 0.4$ for $\De = 1,3$, and 9 respectively. 
Similarly, the local slopes $\zetat$ for the second-order second difference structure functions $\Sdiff ({\rm r})$  are plotted in panel (b). We compute these local exponents for 18 different time-snapshots. The set of 18 exponents for each $r$ is then used to compute the local mean and deviation. We plot the mean values as a solid curve and show the corresponding deviation as shaded regions. The expected value of 3 is marked by a dash-dotted line for reference.

Finally, we show in panel (c) the log-log plots of the integer moments $q$ of the scale-averaged fluid energy dissipation rate $\bra{\epsl^q}$ versus the scale $r$ (for $\De = 3$). The emergence of a clear power-law regime (for $7 \times 10^{-3} \lesssim r/L \lesssim 9 \times 10^{-2})$ enables us to compute the multifractal spectrum (plotted in \Fig{fig:PDFs}  of the main text) using $q \in [-6,6]$.

%
%
%
%

\end{document}